\documentclass[preprints,article,accept,moreauthors]{Definitions/mdpi}
\usepackage{eurosym}
\usepackage{amssymb}

\firstpage{1}
\pubvolume{xx}
\issuenum{1}
\articlenumber{5}
\pubyear{2020}
\copyrightyear{2020}
\history{Received: date; Accepted: date; Published: date}
\Title{Radiation problems accompanying carrier production by an
electric field in the graphene}

\Author{Sergey P. Gavrilov $^{1,3}$\orcidA{}, Dmitry M. Gitman $^{1,2,4}$, Vadim V. Dmitriev$^{5}$, Anatolii D. Panferov$^{5}$\orcidB{} and Stanislav A. Smolyansky$^{1,5}$\orcidC{}*}

\AuthorNames{Firstname Lastname, Firstname Lastname and Firstname Lastname}

\address{%
$^{1}$ \quad Department of Physics, Tomsk State University, 634050 Tomsk, Russia;\\
$^{2}$ \quad P.N. Lebedev Physical Institute, 53 Leninskiy prospect, 119991 Moscow,
Russia;\\
$^{3}$ \quad Department of General and Experimental Physics, Herzen State
Pedagogical University of Russia, Moyka embankment 48, 191186
St.~Petersburg, Russia;\\
$^{4}$ \quad Institute of Physics, University of S\~{a}o Paulo, CP 66318, CEP
05315-970 S\~{a}o Paulo, SP, Brazil;\\
$^{5}$ \quad Department of Physics, Saratov State University, RU-410012 Saratov, Russia.}

\corres{Correspondence: smol@info.sgu.ru}

\abstract{A number of physical processes occurring in a flat one-dimensional graphene structure under the action of strong time-dependent electric fields are considered. It is assumed that the Dirac model can be applied to the graphene as a subsystem of the general system under consideration, which includes an interaction with quantized electromagnetic field. The Dirac model itself in the external electromagnetic field (in particular, the behavior of charged carriers) is treated nonperturbatively with respect to this field within the framework of strong-field QED with unstable vacuum. This treatment is combined with a kinetic description of the radiation of photons from the electron-hole plasma created from the vacuum under the action of the electric field. An interaction with quantized electromagnetic field is described perturbatively. A significant development of the kinetic equation formalism is presented. A number of specific results are derived in course of analytical and numerical study of the equations. We believe that some of predicted effects and properties of considered processes may be verified experimentally. Among these effects, it should be noted a characteristic spectral composition anisotropy of the quantum radiation and a possible presence of even harmonics of the external field in the latter radiation.}

\begin{document}



\section{Introduction}

\label{I}

Particle creation from the vacuum by strong electromagnetic and
gravitational fields is a remarkable quantum effect predicted first in Refs.
\cite{Klein:1929,Klein:1927,Sauter:1931,Sauter:1932,Schwinger:1951} and
studied then in Refs. \cite%
{Nikishov:1970,Nikishov:1970a,Nikishov:1985,Narozhnyi:1970,Narozhnyi:1974}
in the framework of the relativistic quantum mechanics. Its exhaustive
explanations and consistent nonperturbative methods of investigation became
possible in the framework of quantum field theory (QFT). QFT with external
backgrounds is, to a certain extent, an appropriate model for this purpose.
In the framework of such a model, the particle creation is related to a
violation of the vacuum stability. Backgrounds (external fields) that
violate the vacuum stability are electric-like fields that are able to
produce nonzero work when interacting with charged particles. Depending on
the structure of such backgrounds, different approaches for calculating the
effect were proposed and realized. From a quantum mechanical point of view,
the most clear formulation of the problem of particle production from the
vacuum by external fields is possible for time-dependent homogeneous
external electric fields that are switched on and off at infinitely remote
times $t\rightarrow \pm \infty $, respectively.\ Such kind of external
fields are called $t$-electric potential steps ($t$-steps). Scattering,
particle creation from the vacuum and particle annihilation by the $t$-steps
were considered in the framework of the relativistic quantum mechanics, see
Refs. \cite%
{Nikishov:1970,Nikishov:1970a,Nikishov:1985,Narozhnyi:1970,Narozhnyi:1974},
a more complete list of relevant publications can be found in Refs. \cite%
{Ruffini:2010,Gelis:2016}. A general nonperturbative with respect to the
external background formulation of QED was developed in Ref. \cite%
{Gitman:1977,Fradkin:1981,Fradkin:1991}. In contrast to $t$-electric
potential steps, there are many physically interesting situations where the
external backgrounds are constant (time-independent) but spatially
inhomogeneous, for example, concentrated in restricted space areas. The
simplest type of such backgrounds are so-called $x$-electric potential steps
($x$-steps), in which the field is inhomogeneous only in one space direction
and represents a spatial-like potential step for charged particles. The $x$%
-steps can also create particles from the vacuum, the Klein paradox is
closely related to this process \cite%
{Klein:1929,Klein:1927,Sauter:1931,Sauter:1932}. Important calculations of
the particle creation by $x$-steps in the framework of the relativistic
quantum mechanics were presented in Refs. \cite%
{Nikishov:1970,Nikishov:1970a,Nikishov:1985} and developed later in Refs.
\cite{Hansen:1981}. A general nonperturbative with respect to such external
background formulation of QED was formulated in Ref. \cite{Gavrilov:2016}.
It is based on the existence of exact solutions of the Dirac or Klein-Gordon
equation (wave equations, in what follows) with corresponding external
fields. When such solutions can be found and all the calculations can be
done analytically, we refer these examples as exactly solvable cases. Until
now, there are known only few exactly solvable cases related to $t$-steps
and to $x$-steps. In the case of $t$-steps, these are particle creation in a
constant uniform electric field \cite%
{Schwinger:1951,Nikishov:1970,Nikishov:1970a,Nikishov:1985}, in an adiabatic
electric field, \cite{Narozhnyi:1970,Narozhnyi:1974}, in the so-called $T$%
-constant electric field \cite{Bagrov:1975,Gavrilov:1996}, in a periodic
alternating in time electric field \cite%
{Narozhnyi:1970,Narozhnyi:1974,Mostepanenko:1974}, in an exponentially
decaying electric field \cite{Adorno:2015}, in an exponentially growing and
decaying electric fields \cite{Adorno:2016,Adorno:2017} (see Ref. \cite%
{Adorno:2017a} for the review), in a composite electric field \cite%
{Adorno:2018}, and in an inverse-square electric field \cite{Adorno:2018a}.
In the case of $x$-steps these are particle creation in the Sauter electric
field \cite{Gavrilov:2016}, in the so-called $L$-constant electric field
\cite{Gavrilov:2016a}, and in the inhomogeneous exponential peak field \cite%
{Gavrilov:2017} and inverse-square electric field \cite%
{Adorno:2018a,Adorno:2020}.

Until recently, problems related to particle creation from the vacuum had
mostly theoretical interest. This is related to the fact that the vacuum
instability can be observed only in extremely strong external electric
fields of the magnitude of $E_{c}=m^{2}/e\simeq 10^{16}V/cm$ ($E_{c}$ is the
Schwinger's critical field). However, recent technological advances in laser
physics suggest that lasers such as those planned for the Extreme Light
Infrastructure project (ELI) may be able to reach the nonperturbative regime
of pair production in the near future (see review \cite{Dunne:2009}).
Moreover, the situation has changed completely in recent years regarding
applications to condensed matter physics: particle creation became an
observable effect in graphene physics, an area which is currently under
intense development \cite{Castro:2009,Peres:2010,Vozmediano:2010,Das:2011}.
Briefly, this is explained by two facts: first, the low-energy electronic
excitations in the graphene monolayer in the presence of an external
electromagnetic field can be described by the Dirac model \cite%
{Semenoff:1984}, namely, by a $2+1$ quantized Dirac field in such a
background (here, dispersion surfaces are the so-called Dirac cones); and,
second, the gap between the upper and lower branches in the corresponding
Dirac particle spectra is very small, so that the particle creation effect
turns out to be dominant (under certain conditions) as a response to the
applied external electric-like field to the graphene. In particular, such an
effect is crucial for understanding the conductivity in the graphene,
specially in the so-called non-linear regime \cite%
{Allor:2008,Dora:2010,Lewkowicz:2009,Rosenstein:2010,Kao:2010,Lewkowicz:2011}%
. The first experimental observation of non-linear current-voltage
characteristics ($I-V$) of graphene devices and its interpretation in terms
of the pair-creation has been recently reported in \cite{Vandecasteele:2010}%
. In the work \cite{Gavrilov:2012} the quantum electronic and energy
transport in the graphene at low carrier density and low temperatures when
quantum interference effects are important were studied in the framework of
strong field QED. A formulation of the Dirac model\emph{\ }in the Fock space
that includes an interaction of photons with fermions was considered in Ref.
\cite{Gavrilov:2016r}.

Due to a limited number of exactly solvable cases, approaches have been
developed in parallel that make it possible, in the absence of appropriate
exact solutions of the Dirac equation, to use certain approximate methods,
including semiclassical and numerical, for nonperturbative calculations of
quantum effects related to the vacuum instability. In this regard, it should
be noted here the method of quantum kinetic equations (KE). In particular,
this method is well adopted to using numerical calculations. In the
framework of strong field QED, such an approach was also considered in Refs.
\cite{Popov:1973, Grib:1994, Birula:1991,
Schmidt:1998,Kluger:1998,Mamaev:1979} (equivalence of the KE method and
other exact methods was demonstrated in some cases in Refs. \cite%
{Dumlu:2009,Fedotov:2011}) and then applied to problems of QED with strong
external fields (see, for example, \cite{Blaschke:2017}). This approach was
recently adapted to the model of single-layer graphene in \cite%
{Smolyansky:2019,Panferov:2017,Smolyansky:2019a}.

In the present work we propose a generalization of the KE method taking into
account an interaction of carriers with a photon reservoir in the graphene
excited by an uniform, time-dependent electric field. It is assumed that the
interaction of carriers with the external quasiclassical field is taken into
account nonperturbatively, in contrast to their interaction with the
quantized electromagnetic field. Thus, there appear collision integrals (CI)
which take into account interaction with the quantized electromagnetic field
in the single-photon approximation, corresponding to two channels: carrier
redistribution by momenta as a result of a stimulated absorption or emission
of photons and annihilation or creation of pairs. The kinetic description of
these processes required the introduction of an appropriate KE for the
photon subsystem. The Maxwell equations describing the generation of an
internal plasma field close the system of equations.

In what follows, we restrict ourselves by the study of quantum radiation
processes within the framework of the KE approach \cite%
{Blaschke:2011,Smolyansky:2019b}. We consider two different radiation
mechanisms. One of them is a collective mechanism for the generation of
plasma currents and waves (in the spatially uniform case, plasma
oscillations \cite{Kluger:1998,Bloch:1999}). Corresponding electromagnetic
fields can leave the region of active action of the external field and, as a
result, can be detected far from it outside the graphene plane. In the
graphene the excitation region is limited by a simply connected surface and
it is not difficult to find the electromagnetic field in the whole space
from surface currents \cite{Abbott:1985}. A theoretical and experimental
study of this kind of collective radiation was done in Refs. \cite%
{Baudisch:2018,Bowlan:2014} within the framework of an alternative dynamic
approach \cite{Ishikava:2010,Ishikava:2013} and in \cite{Smolyansky:2019} to
the KE approach \cite{Smolyansky:2020b}. In addition to the above mentioned
quasiclassical radiation, a quantum component of the radiation also exists
due to elementary acts of interaction of the carriers with photons. In
graphene, these mechanisms can be taken into account by analogy with the
standard $D=3+1$ QED \cite{Blaschke:2011,Smolyansky:2019b,Smolyansky:2020}.
Outside of the KE approach, this radiation mechanism was studied in Ref.
\cite{Otto:2017}. In the general case, these two mechanisms act
self-consistent: the generation of the photon field leads to a back reaction
problem of the second level, influencing plasma currents and oscillations of
electromagnetic fields.

The approach considered in the present work is based on an adaptation of KE
methods for a description of nonrelativistic plasma-like media in a
nonequilibrium state (see, for example, \cite%
{deGroot:1980,Kadanoff:1962,Zubarev:1996}). The peculiarity of the system
under consideration is that in the presence of external fields, massless
quasiparticle states in the graphene are observed only indirectly and appear
only through macroscopic characteristics, such as currents and the
radiation, which are available for an observation during strongly
nonequilibrium evolution at any time. \textrm{\ }A more detailed discussion
of the KE methods used in the present work and, in particular, the role of
the polarization phenomena is given below in Sect. \ref{V}.

The work is organized as follows. In Subsect. \ref{IIa} we briefly describe
(following in main Refs. \cite{Panferov:2017,
Smolyansky:2019,Smolyansky:2019a,Smolyansky:2020b}) a nonperturbative KE
approach for studying evolution of carrier excitations in monolayer graphene
under the action of an external quasiclassical time dependent electric
field. The set of KE obtained here is an analog of the corresponding
equations which was used in strong field QED (see, e.g. \cite{Blaschke:2017}%
). In the Subsect. \ref{IIb} the KE are generalized to include the
interaction of carriers with the quantized electromagnetic field. They allow
us to consider a set of new problems: the radiation of the quantized field
from the graphene, the photoproduction of carriers under the action of the
quantized electromagnetic field, cascade processes, and so on. The obtained
closed self-consistent system of KE appears as a result of an application of
the truncation procedure of the Bogoliubov--Born--Green--Kirkwood--Yvon
(BBGKY) chains of equations\textbf{\ }in the lower order of the perturbation
theory with respect to the interaction of carriers with the photon
reservoir. In the following Sect. \ref{III} we set the problem how to
consistently treat the radiation of the quantized electromagnetic field in
the physical system under consideration. We argue that for our purposes it
is enough to analyze only photon KE. The corresponding CI are quite complex
functionals containing distribution functions of the carrier and photon
subsystems. Significant simplifications appear in the case of a small photon
density. This case we call the low density approximation. We note that, in
addition, everywhere is used the long wave approximation. Some results
obtained in the framework of such approximations are analyzed in Sect. \ref%
{IV}. Then they are used in Sect. \ref{V} in order to analyze the CI in the
annihilation and the momentum redistribution channels. We also present
parallel numerical estimates of some of the analytical results, and we are
convinced of their qualitative coincidence. In Sect. \ref{VI} we present a
summary of all obtained results, point out open problems, and outlook
possible perspectives.

\section{Kinetic equations describing quantum excitations in graphene placed
in an electric field}

\label{II}

\subsection{Kinetic equations describing zero order processes}

\label{IIa}

As was already mentioned in the Introduction, problems related to the vacuum
instability can be studied in the framework the KE approach. Here we use
such an approach to study the system of electronic excitations (charged
carriers) and their interaction with electromagnetic fields in the graphene
placed in an external electromagnetic field. Conditionally, the
consideration is divided in two steps. On the first step, we consider
production of the charged carriers from the quantum vacuum in the graphene
under the action of a spatially uniform time-dependent external electric
field $\mathbf{E}_{\mathrm{ext}}(t)=\left( E_{\mathrm{ext}}^{1}\left(
t\right) ,E_{\mathrm{ext}}^{2}\left( t\right) \right) $ situated in the
graphene plane $(x^{1},x^{2})$, distracting from a possible interaction of
the carriers with the quantized electromagnetic field and neglecting
possible modification of the external field due to a back reaction.
According to terminology accepted in studying the vacuum instability, on
this first step we consider KE approach for describing only zero order
processes with respect to radiative corrections, see Ref. \cite%
{Gitman:1977,Fradkin:1981,Fradkin:1991}. On the second step, we take into
account the interaction of the carriers with the quantized electromagnetic
field, as well as with the modified due to the back reaction external field
and then, on this base, we study the resulting electromagnetic field in the
graphene. Partially, such a consideration takes into account the first order
processes with respect to the radiative corrections. We stress that the
first step consideration is based on KE approach describing zero order
processes in the graphene placed in an external electric field and is
nonperturbative with respect to the interaction with this field.

The external field is described by electromagnetic potentials $A_{\mathrm{ext%
}}^{\alpha }(t)$, $\alpha =1,2$. In the general case, the carriers in the
graphene are subjected to an effective electric field, \textrm{{\Large \ } }%
\begin{equation}
E^{\alpha }\left( t\right) =\ E_{\mathrm{ext}}^{\alpha }\left( t\right) +E_{%
\mathrm{int}}^{\alpha }\left( t\right) ,\;\alpha =1,2,  \label{2.38}
\end{equation}%
\ given by potentials
\begin{equation}
A^{\alpha }(t)=A_{\mathrm{ext}}^{\alpha }(t)+A_{\mathrm{int}}^{\alpha }(t),
\label{2.2}
\end{equation}%
where the internal field $E_{\mathrm{int}}^{\alpha }\left( t\right) $ is
induced by a back-reaction mechanism (as will be shown, $E_{\mathrm{int}%
}\left( t\right) ${\large \ }can be neglected under some suppositions). We
let the electric field be switched on at $t_{\mathrm{in}}$ and switched off
at $t_{\mathrm{out}}$, so that the interaction between the Dirac field and
the electric field vanishes at all time instants outside the interval $t\in
\left( t_{\mathrm{in}},t_{\mathrm{out}}\right) $.

To describe the carrier quantum motion we use the Dirac model of the
graphene which describes the carries in a vicinity of one of the two Dirac
points at boundaries of the Brillouin zone, see Refs. \cite{Novoselov:2005,
Geim:2007,Castro:2009} for a review. This model considers non interacting
between themselves carriers placed in the external field. A wave function of
a carrier is a two-component spinor $\psi (\mathbf{r},t)$. The latter
satisfies the corresponding massless Dirac equation\footnote{%
Here and what follows $\boldsymbol{\sigma }=(\sigma _{k},\ k=1,2,3)$ are
Pauli matrices,
\[
\sigma _{1}=\begin{pmatrix} 0 & 1 \\ 1 & 0\end{pmatrix},~~\sigma _{2}=%
\begin{pmatrix} 0 & -i \\ i & 0\end{pmatrix},\ \sigma _{3}=%
\begin{pmatrix} 1
& 0 \\ 0 & -1\end{pmatrix}
\]%
and ${\mathrm{v}}_{F}=10^{6}$ m/s is the Fermi velocity.};%
\begin{eqnarray}
&&i\hbar \dot{\psi}(\mathbf{r},t)=h\left( t\right) \psi (\mathbf{r},t),\
h\left( t\right) ={\mathrm{v}}_{F}\mathbf{\hat{P}}\left( t\right)
\boldsymbol{\sigma },\ \mathbf{r}=\left( x^{1},x^{2}\right) ,  \label{2.3} \\
&&\mathbf{\hat{P}}\left( t\right) =\mathbf{\hat{p}}+\frac{e}{c}\mathbf{A}%
(t),\ e>0,\ \mathbf{\hat{p}}=-i\hbar \nabla \ .  \nonumber
\end{eqnarray}%
The wave function $\psi (\mathbf{p},t)$ in the momentum representation is
defined by the decomposition
\begin{equation}
\psi (\mathbf{r},t)=\frac{1}{\sqrt{S}}\sum_{\mathbf{p}}\psi (\mathbf{p}%
,t)e^{i\mathbf{pr}/\hbar },  \label{2.4}
\end{equation}%
where $S$ is the area of the standard box regularization, and satisfies the
following equation:%
\begin{equation}
i\hbar \dot{\psi}\left( \mathbf{p},t\right) =h_{\mathbf{p}}\left( t\right)
\psi \left( \mathbf{p},t\right) ,\;h_{\mathbf{p}}\left( t\right) ={\mathrm{v}%
}_{F}{{\mathbf{P}}}\left( t\right) {{\boldsymbol{\sigma }\ ,}}  \label{2.5}
\end{equation}%
whereas $\mathbf{P}$ is the kinetic momentum,%
\begin{equation}
\mathbf{P}\left( t\right) =\mathbf{p}+\frac{e}{c}\mathbf{A}(t)=\left(
P^{1},P^{2}\right) .  \label{2.6}
\end{equation}

Let us perform an unitary transformation $\psi \left( \mathbf{p},t\right)
=U\left( t\right) \varphi \left( \mathbf{p},t\right) ,$ where the matrix $U$
has the form \cite{Dora:2010}:
\begin{equation}
U(t)=\frac{1}{\sqrt{2}}\left(
\begin{array}{cc}
\exp (-i\varkappa /2) & \exp (-i\varkappa /2) \\
\exp (i\varkappa /2) & -\exp (i\varkappa /2)%
\end{array}%
\right) .  \label{2.7}
\end{equation}%
Then we come to an auxiliary quasienergy eigenvalue problem for the
transformed Hamiltonian $\tilde{h}_{\mathbf{p}}(t)=U^{\dag }(t)h_{\mathbf{p}%
}\left( t\right) U(t).$ We fix the parameter $\varkappa $ by the condition $%
\tan \varkappa =P^{2}/P^{1}$ such that the corresponding quasienergy (the
excitation quasienergy or a dispersion law) is $\varepsilon (\mathbf{p},t)=%
\mathrm{v}_{F}\sqrt{\mathbf{P}^{2}}$ and:
\begin{equation}
\tilde{h}_{\mathbf{p}}(t)u_{\pm 1}=\pm \varepsilon (\mathbf{p},t)u_{\pm 1},\
u_{+1}=\left(
\begin{array}{c}
1 \\
0%
\end{array}%
\right) ,\ u_{-1}=\left(
\begin{array}{c}
0 \\
1%
\end{array}%
\right) .  \label{2.8}
\end{equation}%
The spinor $\varphi \left( \mathbf{p},t\right) $ satisfies the equation
\begin{equation}
i\hbar \dot{\varphi}(\mathbf{p},t)=\tilde{h}_{\mathbf{p}}(t)\varphi (\mathbf{%
p},t)+\frac{1}{2}\lambda \hbar \sigma _{1}\varphi (\mathbf{p},t),
\label{2.9}
\end{equation}%
where the excitation function $\lambda (\mathbf{p},t)$ is determined from
the equation $2iU^{\dagger }\dot{U}=\lambda \sigma _{1}$ and has the form:
\begin{equation}
\lambda (\mathbf{p},t)=\frac{ev_{F}^{2}[E_{\mathrm{ext}}^{1}(t)P^{2}-E_{%
\mathrm{ext}}^{2}(t)P^{1}]}{\varepsilon ^{2}(\mathbf{p},t)}.  \label{2.10}
\end{equation}

Note that the Dirac Hamiltonians $h\left( t\right) $ do not commute at
distinct time instants, $\left[ \tilde{h}_{\mathbf{p}}\left( t\right) ,%
\tilde{h}_{\mathbf{p}}\left( t^{\prime }\right) \right] \neq 0$ if $t\neq
t^{\prime }$. In the model under consideration, the dispersion law holds
true in a vicinity of the Dirac point $\mathbf{P}^{2}=0$ at the boundaries
of the Brillouin zone. This model corresponds to low-energy excitations of
the carries. However, the approach under consideration allows a
generalization to a tight-binding model of the nearest neighbor interaction
(see Refs. \cite{Castro:2009,Kao:2010,Gusynin:2007}) as was demonstrated in
Ref. \cite{Smolyansky:2019}.

In which follows we are going to consider the so-called adiabatic ansatz%
{\large \ }(alternatively called the quasiparticle representation) which is
widely used in semiclassical approximations and numeric calculations, see,
e.g., Refs. \cite%
{Popov:1973,Birula:1991,Schmidt:1998,Kluger:1998,Dumlu:2009,Fedotov:2011,Dabrowski:2014}%
.

We recall that there are two species of fermions in the model, corresponding
to excitations about two distinct Dirac points in the Brillouin zone, i.e.,
each of species belongs to a distinct valley. The algebra of $\gamma $%
-matrices has two inequivalent representations in $(2+1)$-dimensions and a
distinct (pseudospin) representation is associated with each Dirac point.
Another doubling of fields is due to the (real) spin of the electron. As a
result, there are four species of fermions in the model. Thus, in order to
find real mean values of physical quantities, one should to take into
account the degeneracy factor $N_{f}=4$. We note that a transition to the
quasiparticle representation in the model was used, for example, in Refs.
\cite{Smolyansky:2019,Klimchitskaya:2013, Dora:2010,Smolyansky:2020b}. Below
we follow the works by \cite{Smolyansky:2019,Smolyansky:2020b}.

A quantum Dirac field ${\Psi (\mathbf{r},t)}$ associated with the function $%
\psi (\mathbf{r},t)$ satisfies Eq. (\ref{2.3}) and the standard equal-time
canonical anticommutation relations. It describes a fermion species of the
model. The field operator $\Psi (\mathbf{p},t)$ in the momentum
representation is defined by the decomposition:
\begin{equation}
\Psi (\mathbf{r},t)=\frac{1}{\sqrt{S}}\sum_{\mathbf{p}}\Psi (\mathbf{p}%
,t)e^{i\mathbf{pr}/\hbar }\ .  \label{2.11}
\end{equation}%
Then it is convenient to introduce the field operator $\Phi (\mathbf{p}%
,t)=U^{-1}\left( t\right) \Psi \left( \mathbf{p},t\right) $, which satisfies
equation (\ref{2.9}). According to the adiabatic ansatz, we define two kinds
of creation and annihilation operators ($a^{\dagger }(\mathbf{p},t),a(%
\mathbf{p},t)$ and $b^{\dagger }(\mathbf{p},t),b(\mathbf{p},t)$) decomposing
the operator $\Phi \left( \mathbf{p},t\right) $ into solutions $u_{\pm 1}$
(see Eq. (\ref{2.8})),
\begin{equation}
\Phi (\mathbf{p},t)=a(\mathbf{p},t)u_{+1}+b^{\dagger }(-\mathbf{p},t)u_{-1}\
.  \label{2.12}
\end{equation}%
Their nonzero anticommutation relations read:%
\begin{equation}
\left[ a(\mathbf{p},t),a^{\dagger }(\mathbf{p}^{\prime },t)\right] _{+}=%
\left[ b(\mathbf{p},t),b^{\dagger }(\mathbf{p}^{\prime },t)\right]
_{+}=\delta _{\mathbf{p,p}^{\prime }}\ .  \label{2.13}
\end{equation}%
In each time instant $t$, one can formally introduce a Fock space equipped
by instantaneous vacuum vectors $|0,t\rangle ,$%
\begin{equation}
a(\mathbf{p},t)|0,t\rangle =b(\mathbf{p},t)|0,t\rangle =0,\ \forall \mathbf{%
p,}  \label{2.14}
\end{equation}%
and a corresponding basis originated by the action of the creation operators
$a^{\dagger }(\mathbf{p},t)$ and $b^{\dagger }(\mathbf{p},t)$ on the
corresponding vacuum vectors. Eq. (\ref{2.9}) for the operator $\Phi \left(
\mathbf{p},t\right) $ implies the following equations for the creation and
annihilation operators:%
\begin{eqnarray}
&&i\hbar \dot{a}(\mathbf{p},t)=\varepsilon (\mathbf{p},t)a(\mathbf{p},t)-%
\frac{1}{2}\hbar \lambda (\mathbf{p},t)b^{\dagger }(-\mathbf{p},t),
\nonumber \\
&&i\hbar \dot{b}(-\mathbf{p},t)=\varepsilon (\mathbf{p},t)b(-\mathbf{p},t)+%
\frac{1}{2}\hbar \lambda (\mathbf{p},t)a^{\dagger }(-\mathbf{p},t).
\label{2.15}
\end{eqnarray}

The QFT Hamiltonian and the corresponding charge operator $Q$ in the model
have the form:
\begin{eqnarray}
&&H(t)=\int {d\mathbf{r}\Psi ^{\dagger }(\mathbf{r},t)h\left( t\right) \Psi (%
\mathbf{r},t)\ ,}  \nonumber \\
&&Q(t)=-\frac{e}{2}\int {d\mathbf{r}}\left[ {\Psi ^{\dagger }(\mathbf{r}%
,t),\Psi (\mathbf{r},t)}\right] ,\ {}  \label{2.16}
\end{eqnarray}%
where the integration is over the finite area $S$. They can be diagonalized
at any time instant $t$ using decomposition (\ref{2.12}),%
\begin{eqnarray}
&&H(t)=\sum_{\mathbf{p}}\varepsilon (\mathbf{p},t)\left[ a^{\dagger }(%
\mathbf{p},t)a(\mathbf{p},t)-b(-\mathbf{p},t)b^{\dagger }(-\mathbf{p},t)%
\right] ,  \nonumber \\
&&Q=-e\sum_{\mathbf{p}}\left[ a^{\dagger }(\mathbf{p},t)a(\mathbf{p}%
,t)-b^{\dagger }(-\mathbf{p},t)b(-\mathbf{p},t)\right] .  \label{2.17}
\end{eqnarray}

The introduced above creation and annihilation operators become \textrm{in}-
and\textrm{\ out}-operators of real quasiparticle at time instants $t_{%
\mathrm{in}}$ and $t_{\mathrm{out}}$ because the external electric field
vanishes for $t\in \left( -\infty ,t_{\mathrm{in}}\right) \cup \left( t_{%
\mathrm{out}},+\infty \right) .$ They act in the corresponding Fock spaces
with initial and final vacua $|0,\mathrm{in}\rangle =|0,t_{\mathrm{in}%
}\rangle $ and $|0,\mathrm{out}\rangle =|0,t_{\mathrm{out}}\rangle $
respectively. One interprets $a^{\dagger }(\mathbf{p},t_{\mathrm{in}})$ and $%
a(\mathbf{p},t_{\mathrm{in}})$ as creation and annihilation operators of
initial electrons, $b^{\dagger }(\mathbf{p},t_{\mathrm{in}})$ and $b(\mathbf{%
p},t_{\mathrm{in}})$ as the creation and annihilation operators of initial
holes, whereas $a^{\dagger }(\mathbf{p},t_{\mathrm{in}})a(\mathbf{p},t_{%
\mathrm{in}})$ and $b^{\dagger }(\mathbf{p},t_{\mathrm{in}})b(\mathbf{p},t_{%
\mathrm{in}})$ are operators of initial electron and hole numbers,
respectively. Operators $a^{\dagger }(\mathbf{p},t_{\mathrm{out}})$ and $a(%
\mathbf{p},t_{\mathrm{out}})$ are interpreted as creation and annihilation
operators of final electrons, $b^{\dagger }(\mathbf{p},t_{\mathrm{out}})$
and $b(\mathbf{p},t_{\mathrm{out}})$ as creation and annihilation operators
of final holes, whereas $a^{\dagger }(\mathbf{p},t_{\mathrm{out}})a(\mathbf{p%
},t_{\mathrm{out}})$ and $b^{\dagger }(\mathbf{p},t_{\mathrm{out}})b(\mathbf{%
p},t_{\mathrm{out}})$ are interpreted as operators of final electron and
hole numbers, respectively.

Since the particles are massless in the model under consideration, and spin
degrees of freedom are hidden and manifest themselves only in the population
of states, the charge is the only characteristic of the quasiparticles. It
can be shown that the equation $\left[ Q(t),H(t)\right] =0$ holds true at
any time moment and, therefore, the electroneutrality of the system is
preserved over time.

It is useful to introduce the following auxiliary distribution functions of
quasiparticles (time-evolving adiabatic particle numbers)
\begin{eqnarray}
f^{e}(\mathbf{p},t) &=&\langle 0,\mathrm{in}|a^{+}(\mathbf{p},t)a(\mathbf{p}%
,t)|0,\mathrm{in}\rangle ,  \label{2.18} \\
f^{h}(\mathbf{p},t) &=&\langle 0,\mathrm{in}|b^{+}(-\mathbf{p},t)b(-\mathbf{p%
},t)|0,\mathrm{in}\rangle .  \label{2.19}
\end{eqnarray}%
Averaging charge operator (\ref{2.16}) over the \textrm{in}-vacuum, and
taking into account the charge conservation low, we obtain
\begin{equation}
f^{e}(\mathbf{p},t)=f^{h}(\mathbf{p},t)=f(\mathbf{p},t)\,.  \label{2.20}
\end{equation}%
Since $a(\mathbf{p},t_{\mathrm{in}})|0,\mathrm{in}\rangle =b(\mathbf{p},t_{%
\mathrm{in}})|0,\mathrm{in}\rangle =0,$ the initial value of the function $f(%
\mathbf{p},t)$ is zero, $f(\mathbf{p},t_{\mathrm{in}})=0$. When the electric
field is turned off in an asymptotically distant future, the dispersion laws
of quasiparticles goes to the mass surface,%
\[
\varepsilon _{\mathrm{out}}(\mathbf{p},t)=\mathrm{v}_{F}\sqrt{\left( \mathbf{%
p+}\frac{e}{c}\mathbf{A}_{\mathrm{out}}\right) ^{2}},\ \mathbf{A}_{\mathrm{%
out}}=\lim_{t\rightarrow \infty }\mathbf{A}\left( t\right) \ .
\]

Then functions (\ref{2.18}) and (\ref{2.19}) describe momentum distributions
of real (observable) particles,%
\begin{equation}
f(\mathbf{p},t_{\mathrm{out}})=\langle 0,\mathrm{in}|a^{+}(\mathbf{p},t_{%
\mathrm{out}})a(\mathbf{p},t_{\mathrm{out}})|0,\mathrm{in}\rangle =\langle 0,%
\mathrm{in}|b^{+}(\mathbf{p},t_{\mathrm{out}})b(\mathbf{p},t_{\mathrm{out}%
})|0,\mathrm{in}\rangle .  \label{2.21}
\end{equation}

To get a closed set of KE, we differentiate $f(\mathbf{p},t)$ with respect
to the time. Then using Eqs. (\ref{2.15}) we obtain:
\begin{equation}
\dot{f}(\mathbf{p},t)=\frac{i\lambda \left( \mathbf{p},t\right) }{2}\left[
f^{(+)}(\mathbf{p},t)-f^{(-)}(\mathbf{p},t)\right] ,  \label{2.22}
\end{equation}%
where the following anomalous expectation values
\begin{eqnarray}
&&f^{(+)}(\mathbf{p},t)=\langle 0,\mathrm{in}|a^{+}(\mathbf{p},t)b^{+}(-%
\mathbf{p},t)|0,\mathrm{in}\rangle ,  \nonumber \\
&&f^{(-)}(\mathbf{p},t)=\langle 0,\mathrm{in}|b(-\mathbf{p},t)a(\mathbf{p}%
,t)|0,\mathrm{in}\rangle ~  \label{2.23}
\end{eqnarray}%
are introduced and $\lambda \left( \mathbf{p},t\right) $ is given by Eq. (%
\ref{2.10}).

Time derivatives of the functions $f^{(\pm )}(\mathbf{p},t)$ have the form:%
\begin{eqnarray}
\dot{f}^{(+)}(\mathbf{p},t) &=&\frac{2i}{\hbar }\varepsilon (\mathbf{p}%
,t)f^{(+)}(\mathbf{p},t)-\frac{i\lambda (\mathbf{p},t)}{2}[1-2f(\mathbf{p}%
,t)],  \nonumber \\
\dot{f}^{(-)}(\mathbf{p},t) &=&\frac{-2i}{\hbar }\varepsilon (\mathbf{p}%
,t)f^{(-)}(\mathbf{p},t)+\frac{i\lambda (\mathbf{p},t)}{2}[1-2f(\mathbf{p}%
,t)].  \label{2.24}
\end{eqnarray}

As a result of the integration of these equations over time and substitution
into Eq. (\ref{2.22}), we obtain a KE of non-Markovian type in the following
form:%
\begin{eqnarray}
&&\dot{f}(\mathbf{p},t)=I(\mathbf{p},t),  \nonumber \\
&&I(\mathbf{p},t)=\frac{1}{2}\lambda (\mathbf{p},t)\int_{t_{0}}^{t}\lambda (%
\mathbf{p},t^{\prime })[1-2f(\mathbf{p},t^{\prime })]\cos \theta (\mathbf{p}%
;t,t^{\prime })dt^{\prime },  \label{2.25a}
\end{eqnarray}%
where the phase\textbf{\ }$\theta $ in the source function $I(\mathbf{p},t)$
reads:%
\begin{equation}
\theta (\mathbf{p};t,t^{\prime })=\frac{2}{\hbar }\int_{t^{\prime
}}^{t}dt^{\prime \prime }\varepsilon (\mathbf{p},t^{\prime \prime }).
\label{2.27}
\end{equation}

The main task of the first stage is to find the distribution function $%
f\left( \mathbf{p},t\right) $ of created carriers. To this end it is
convenient to introduce new functions $u(\mathbf{p},t)$ and $\nu (\mathbf{p}%
,t)$ and reduce integro-differential Eq. (\ref{2.25a}) to an equivalent set
of ordinary differential equations:%
\begin{eqnarray}
&&\dot{f}(\mathbf{p},t)=\frac{1}{2}\lambda (\mathbf{p},t)u(\mathbf{p},t),
\nonumber \\
&&\dot{u}(\mathbf{p},t)=\lambda (\mathbf{p},t)\left[ 1-2f(\mathbf{p},t)%
\right] -\frac{2\varepsilon (\mathbf{p},t)}{\hbar }\nu (\mathbf{p},t),\ \dot{%
\nu}(\mathbf{p},t)=\frac{2\varepsilon (\mathbf{p},t)}{\hbar }u(\mathbf{p},t).
\label{2.28}
\end{eqnarray}%
The functions $u(\mathbf{p},t)$ and $\nu (\mathbf{p},t)$ describe
polarization effects and are expressed in terms of anomalous means (\ref%
{2.23}) as:
\begin{equation}
u(\mathbf{p},t)=f^{(+)}(\mathbf{p},t)+f^{(-)}(\mathbf{p},t),\ \nu (\mathbf{p}%
,t)=i\left[ f^{(+)}(\mathbf{p},t)-f^{(-)}(\mathbf{p},t)\right] .
\label{2.29}
\end{equation}

The total Hamiltonian of the fermion subsystem $H_{\mathrm{tot}}(t)=H(t)+H_{%
\mathrm{pol}}(t)$ contains two parts: the Hamiltonian $H(t)$ of the
quasiparticle excitations (\ref{2.17}) and a polarization Hamiltonian $H_{%
\mathrm{pol}}(t)$ which corresponds to the second term in the RHS of Eq. (%
\ref{2.9}),%
\begin{equation}
H_{\mathrm{pol}}(t)=-i\frac{\hbar }{2}\sum_{\mathbf{p}}\lambda (\mathbf{p},t)%
\left[ a^{\dagger }(\mathbf{p},t)b^{\dagger }(-\mathbf{p},t)-b(-\mathbf{p}%
,t)a(\mathbf{p},t)\right] .  \label{2.30}
\end{equation}

The corresponding vacuum polarization energy density can be represented as
\begin{equation}
E_{\mathrm{pol}}(t)=-\frac{\hbar }{2S}\sum_{\mathbf{p}}\lambda (\mathbf{p}%
,t)v(\mathbf{p},t),  \label{2.31}
\end{equation}%
where $v(\mathbf{p},t)$is given by Eq. (\ref{2.29}).

The vacuum mean value of the current density%
\[
\mathbf{j}(\mathbf{r},t)=\langle 0,\mathrm{in}|e{\mathrm{v}}_{F}\Psi
^{\dagger }(\mathbf{r},t){\boldsymbol{\sigma }}\Psi (\mathbf{r},t)|0,\mathrm{%
in}\rangle
\]%
can be written in the following form:
\begin{equation}
\mathbf{j}\left( t\right) =\mathbf{j}_{\mathrm{cond}}\left( t\right) +%
\mathbf{j}_{\mathrm{pol}}\left( t\right) ,  \label{2.33}
\end{equation}%
see \cite{Smolyansky:2019,Smolyansky:2020b}. Here the conductivity\textbf{\ }%
$\mathbf{j}_{\mathrm{cond}}\left( t\right) $ and polarization current\textbf{%
\ }$\mathbf{j}_{\mathrm{pol}}\left( t\right) $ densities are\textbf{:}
\begin{eqnarray}
j_{\mathrm{cond}}^{\alpha }\left( t\right) &=&\frac{2e}{S}\sum_{\mathbf{p}%
}v^{\alpha }(\mathbf{p},t)f(\mathbf{p},t)\ ,  \label{2.34} \\
j_{\mathrm{pol}}^{\alpha }\left( t\right) &=&-\frac{e}{S}\sum_{\mathbf{p}}%
\tilde{v}^{\alpha }(\mathbf{p},t)u(\mathbf{p},t)\ ,  \label{2.35}
\end{eqnarray}%
where
\begin{equation}
v^{\alpha }(\mathbf{p},t)=\frac{\partial \varepsilon \left( \mathbf{p}%
,t\right) }{\partial P^{\alpha }}=\frac{{\mathrm{v}}_{F}^{2}P^{\alpha }}{%
\varepsilon ^{2}\left( \mathbf{p},t\right) }  \label{2.36}
\end{equation}%
is a group velocity and $\tilde{v}^{\alpha }(\mathbf{p},t)$ is the vector of
the so-called conjugate velocity that is determined by components of vector (%
\ref{2.36}) as follows: $\tilde{v}^{\alpha }(\mathbf{p},t)=\left(
v^{2},-v^{1}\right) $. Note that the total current of all fermion species is%
{\Large \ }$N_{f}\mathbf{j}\left( t\right) $. The role of current densities (%
\ref{2.34}) and (\ref{2.35}) in the electromagnetic emission in the graphene
is discussed in Sect. \ref{V}.

The plasma classical electric field $\mathbf{E}_{\mathrm{int}}(t)$ is
generated by the internal current $N_{f}\mathbf{j}\left( t\right) $\emph{\ }%
and satisfies the Maxwell equation,{\Large \ }%
\begin{equation}
\mathbf{\dot{E}}_{\mathrm{int}}(t)=-N_{f}\mathbf{j}\left( t\right) .
\label{2.37}
\end{equation}%
This field contributes to the effective quasiclassical electric field (\ref%
{2.38}) that in its turn has effect on the dynamics of carriers according to
Eq. (\ref{2.28}).

The above formulation of the quantum kinetic theory describing excitations
in the graphene \cite{Smolyansky:2019,Smolyansky:2020b} is constructed by
analogy with the quantum kinetic theory describing the vacuum production of $%
e^{-}e^{+}$ plasma (see, e.g., \cite{Grib:1994,Blaschke:2009}).

In the thermodynamical limit $S\rightarrow \infty $, replacing the sum over
the momenta in Eqs.~(\ref{2.34}) and (\ref{2.35}) by an integral the
conductivity and polarization current densities take the form\textrm{\ }
\begin{eqnarray}
&&\mathbf{j}_{\mathrm{cond}}\left( t\right) =\frac{2e}{(2\pi \hbar )^{2}}%
\int \mathbf{v}(\mathbf{p},t)f(\mathbf{p},t)d\mathbf{p},  \label{2.39} \\
&&\mathbf{j}_{\mathrm{pol}}\left( t\right) =-\frac{e}{(2\pi \hbar )^{2}}\int
\varepsilon (\mathbf{p},t)\mathbf{l}(\mathbf{p},t)u(\mathbf{p},t)d\mathbf{p},
\label{2.40}
\end{eqnarray}%
where $\mathbf{v}(\mathbf{p},t)$ given by Eq. (\ref{2.36}) is a propagation
velocity of quasiparticle excitations, $\mathbf{l}(\mathbf{p},t)$ is a
polarization function,%
\begin{eqnarray}
&&e\mathbf{l}(\mathbf{p},t)=(\partial \lambda (\mathbf{p},t)/\partial E_{%
\mathrm{ext}}^{1}\left( t\right) ,\ \partial \lambda (\mathbf{p},t)/\partial
E_{\mathrm{ext}}^{2}\left( t\right) ),  \nonumber \\
&&l_{1}(\mathbf{p},t)=v_{F}^{2}P_{2}/\varepsilon ^{2}(\mathbf{p},t),\ l_{2}(%
\mathbf{p},t)=-v_{F}^{2}P_{1}/\varepsilon ^{2}(\mathbf{p},t).  \label{2.41}
\end{eqnarray}%
The function $\lambda (\mathbf{p},t)$ (\ref{2.10}) can be expressed via $%
\mathbf{l}(\mathbf{p},t)$ as follows:%
\begin{equation}
\lambda (\mathbf{p},t)=e\mathbf{E}_{\mathrm{ext}}(t)\mathbf{l}(\mathbf{p},t).
\label{2.42}
\end{equation}

Thus, we see that polarization energy (\ref{2.31}) and polarization current (%
\ref{2.40}) are expressed in terms of polarization functions (\ref{2.29}).
We note that the separation of the total current in the sum of the
conduction current and the polarization current is in a certain sense
conditional but useful in approximate calculations.

Among the macroscopic averages it is necessary to refer also the number
density of pairs of $N_{f}$ fermion species%
\begin{equation}
n\left( t\right) =\frac{N_{f}}{(2\pi \hbar )^{2}}\int f(\mathbf{p},t)d%
\mathbf{p},  \label{dop1}
\end{equation}%
Following the works \cite{Grib:1994,Mamaev:1979}, we can discover that for a
finite $t$\ the asymptotic behavior of solutions of system (\ref{2.28}) for $%
\varepsilon (p,t)\rightarrow \infty $\ is described by the following leading
terms{\large \ }%
\begin{equation}
f(\mathbf{p},t)\approx \frac{1}{16}\left( \frac{\lambda (\mathbf{p},t)}{%
\varepsilon (\mathbf{p},t)}\right) ^{2},\;u(\mathbf{p},t)\approx \frac{1}{%
4\varepsilon (\mathbf{p},t)}\frac{d}{dt}\frac{\lambda (\mathbf{p},t)}{%
\varepsilon (\mathbf{p},t)},\;\nu (\mathbf{p},t)\approx \frac{\lambda (%
\mathbf{p},t)}{2\varepsilon (\mathbf{p},t)}\ .  \label{dop2}
\end{equation}%
Therefore, integrals (\ref{2.39}), (\ref{2.40}), and (\ref{dop1}) converge.
Note that for $t\rightarrow \infty $\ the term (\ref{2.40}) containing $%
u(p,t)$\ vanishes because it is represented via an integral over momenta of
a rapidly oscillating function.{\large \ }Assuming the external field being
switched off for $t\rightarrow \infty $\ the term $f(p,t)$\ exponentially
disappears with growth of $\varepsilon (p,t)${\large .}

\subsection{Inclusion an interaction with quantized electromagnetic field}

\label{IIb}

Here we assume that excitations in the graphene interact both with effective
classical electromagnetic field (\ref{2.2}) and with the quantized
electromagnetic field situated in the graphene plane. Below, we generalize
the KE for such an extended system. The total QFT Hamiltonian $H_{\mathrm{tot%
}}(t)$ that corresponds to the system consists of a part which is originated
from the Dirac model of excitations in the graphene interacting both with an
external classical electromagnetic field and with the quantized
electromagnetic field and of a Hamiltonian of the quantized electromagnetic
field. Note that we consider explicitly an interaction with only one of a
fermion species of the model.{\Large \ }Thus, to find real mean values of
physical quantities, one should to take into account the degeneracy factor%
{\Large \ }$N_{f}=4${\Large . }

In the quasiparticle representation, introduced in Sect. \ref{IIa}, such a
Hamiltonian reads:%
\begin{equation}
H_{\mathrm{tot}}(t)=H(t)+H_{\mathrm{pol}}(t)+H_{\mathrm{int}}(t)+H_{\mathrm{A%
}}(t),  \label{2.43}
\end{equation}%
where Hamiltonians $H(t)$, given by Eq. (\ref{2.17}), and $H_{\mathrm{pol}%
}(t)$ given by Eq. (\ref{2.29}) describe excitations in the graphene and
their interaction with existing in the graphene quasiclassical
electromagnetic field $\mathbf{E}_{\mathrm{ext}}(t)$ whereas $H_{\mathrm{A}%
}(t)$ is a Hamiltonian of the free quantized electromagnetic field in the
Coulomb gauge.

The electromagnetic field is not confined to the graphene surface, $z=0$,
but rather propagates in the ambient $3+1$\ dimensional space-time, where $z$%
\ is the coordinate of axis normal to the graphene plane. We allow the
graphene sheet to have a global momentum $p_{z}$ along the $z$ axis, in
order to account for the possibility of a momentum transfer in this
direction to some external system. However, only the projection of the
electromagnetic field operator on the graphene plane $\mathbf{A}(x)=\left(
A^{1}(x),A^{2}(x)\right) $ interacts with electrons and holes.\emph{\ }Thus,
we can exclude the noninteracting component, $A^{3}(x)$ of the
electromagnetic field operator from the consideration and which corresponds
to the case $z=0$\emph{. }If the \ wave vector of a photon field, $\left(
k^{1},k^{2},k_{z}\right) $, has component $k_{z}\neq 0$ then the photon
immediately leaves the interaction area. Such a free photon field can be
represented with standard annihilation and creation operators and formulate
in such terms a theory of emission in the first-order approximation with
respect of electron-photon interaction \cite{Gavrilov:2016r}.\emph{\ }%
However, in the present model we are interested in a different emission
mechanism, which allows one to consider the photon field propagating along
the graphene plane and having a sufficiently small momenta\emph{\ }$%
k_{z}\rightarrow 0$. We chose the range of $k_{z}$\ so as to emitted photon
did not leave a graphene film of thickness\emph{\ }$d$\emph{. }Such a
condition determines a limiting angle between the emission direction and the
graphene plane,\emph{\ }%
\begin{equation}
\left\vert k_{z}\right\vert /k<\varphi ,\;\varphi \sim d/\sqrt{S}\ll 1,
\label{gav1}
\end{equation}%
where $k=|\mathbf{k}|$ and $\mathbf{k}=\left( k^{1},k^{2}\right) $ is a
two-component wave vector.

The operator of such a field can be written in the following form:
\begin{eqnarray}
&&\hat{A}^{\alpha }(\mathbf{r},t)=\hat{A}^{(+)\alpha }(\mathbf{r},t)+\hat{A}%
^{(-)\alpha }(\mathbf{r},t),\ \alpha =1,2\ ,  \nonumber \\
&&\hat{A}^{(\pm )\alpha }(\mathbf{r},t)=\sqrt{\frac{\hbar c}{V}}\sum_{%
\mathbf{k}}\sum_{k_{z}<\varphi k}\frac{1}{\sqrt{2k}}\hat{A}^{(\pm )\alpha
}(\pm \mathbf{k},k_{z},t)e^{-i\mathbf{kr}}\ ,  \nonumber \\
&&k=|\mathbf{k}|,\ \left[ \hat{A}^{(\pm )\alpha }(\mathbf{k},k_{z},t)\right]
^{\dag }=A^{(\mp )\alpha }(-\mathbf{k},-k_{z},t)\ ,  \label{2.44}
\end{eqnarray}%
where \emph{\ }$V=SL$\emph{\ }is the volume of the regularization box ($L$\
is the length of the edge of the box normal to the graphene plane). This
field is polarized and its polarization vector is transversal to the
wavevector $\mathbf{k}$. The operators $\hat{A}^{(\pm )\alpha }(\mathbf{k}%
,k_{z},t)$ satisfy standard equal time commutation relations:
\begin{eqnarray}
\left[ A^{(-)\alpha }(\mathbf{k},k_{z},t),A^{(+)\beta }(\mathbf{k}^{\prime
},k_{z}^{\prime },t)\right] &=&\delta _{\alpha \beta }\delta _{\mathbf{kk}%
^{\prime }}\delta _{k_{z},k_{z}^{\prime }}\ ,  \nonumber \\
\left[ A^{(\pm )\alpha }(\mathbf{k},k_{z},t),A^{(\pm )\beta }(\mathbf{k}%
^{\prime },k_{z}^{\prime },t)\right] &=&0\ .  \label{2.45}
\end{eqnarray}%
Then, the Hamiltonian $H_{\mathrm{A}}(t)$ of the free quantized
electromagnetic field can be written as
\begin{equation}
H_{\mathrm{A}}(t)=c\hbar \sum_{\mathbf{k}}\sum_{k_{z}<\varphi k}k\hat{A}%
^{(+)\alpha }(\mathbf{k},k_{z},t)\hat{A}^{(-)\alpha }(\mathbf{k},k_{z},t),
\label{2.46}
\end{equation}

The effective interaction of the carriers in graphene with the quantized
electromagnetic field can be defined as:\textbf{\ }
\begin{equation}
H_{\mathrm{int}}(t)=-e\frac{v_{F}}{c}\int d\mathbf{r}\Psi ^{\dag }(\mathbf{r}%
,t){\boldsymbol{\sigma }}\Psi (\mathbf{r},t)\hat{\mathbf{A}}(\mathbf{r},t).
\label{2.48}
\end{equation}%
Substituting decompositions (\ref{2.11}), (\ref{2.12}) and (\ref{2.44}) in
Hamiltonian (\ref{2.48}), we obtain:
\begin{eqnarray}
&&\ H_{\mathrm{int}}(t)=-ev_{F}\sum_{\mathbf{p},\mathbf{k}%
}\sum_{k_{z}<\varphi k}\sqrt{\frac{\hbar }{2cVk}}\left\{ \Gamma
_{uu}^{\alpha }(\mathbf{p},\mathbf{p}-\hbar \mathbf{k};t)a^{\dag }(\mathbf{p}%
,t)a(\mathbf{p}-\hbar \mathbf{k},t)\right.  \nonumber \\
&&+\Gamma _{uv}^{\alpha }(\mathbf{p},\mathbf{p}-\hbar \mathbf{k};t)a^{\dag }(%
\mathbf{p},t)b^{\dag }(-\mathbf{p}+\hbar \mathbf{k},t)+\Gamma _{vu}^{\alpha
}(\mathbf{p},\mathbf{p}-\hbar \mathbf{k};t)b(-\mathbf{p},t)a(\mathbf{p}%
-\hbar \mathbf{k},t)  \nonumber \\
&&+\left. \Gamma _{vv}^{\alpha }(\mathbf{p},\mathbf{p}-\hbar \mathbf{k};t)b(-%
\mathbf{p},t)b^{\dag }(-\mathbf{p}+\hbar \mathbf{k},t)\right\} \hat{A}%
^{\alpha }(\mathbf{k},k_{z},t).  \label{2.49}
\end{eqnarray}%
Here $\Gamma _{\xi \eta }^{\alpha }(\mathbf{p},\mathbf{p}^{\prime },t)$ are
vertex matrix functions,
\begin{equation}
\Gamma _{\xi \eta }^{\alpha }(\mathbf{p},\mathbf{p}^{\prime },t)=\xi ^{\dag
}U^{\dag }(\mathbf{p},t)\sigma ^{\alpha }U(\mathbf{p}^{\prime },t)\eta ,\
\Gamma _{\xi \eta }^{\alpha \ast }(\mathbf{p},\mathbf{p}^{\prime };t)=\Gamma
_{\eta \xi }^{\alpha }(\mathbf{p}^{\prime },\mathbf{p};t),  \label{2.50}
\end{equation}%
spinors $\xi $ and $\eta $ are given by Eqs. (\ref{2.8}), and the evolution
matrix $U(\mathbf{p},t),$ which describes the influence of the external
field on the interaction of the introduced quasiparticles with photons is
given by Eq. (\ref{2.7}).

Hamiltonian (\ref{2.43}), being written in the quasiparticle representation,
determines the time evolution of creation and annihilation operators of all
the particles,

\begin{eqnarray}
&&\dot{a}(\mathbf{p},t)=-\frac{i}{\hbar }\varepsilon (\mathbf{p},t)a(\mathbf{%
p},t)+\frac{i}{2}\lambda (\mathbf{p},t)b^{\dag }(-\mathbf{p},t)+ev_{F}\sum_{%
\mathbf{k}}\sum_{k_{z}<\varphi k}\frac{i}{\sqrt{2\hbar cVk}}  \nonumber \\
&&\times \left\{ \Gamma _{uu}^{\alpha }(\mathbf{p},\mathbf{p}^{\prime };t)a(%
\mathbf{p}-\hbar K,t)+\Gamma _{uv}^{\alpha }(\mathbf{p},\mathbf{p}^{\prime
};t)b^{\dag }(-\mathbf{p}^{\prime },t)\right\} \hat{A}^{\alpha }(\mathbf{k}%
,k_{z},t),  \nonumber \\
&&\dot{b}(-\mathbf{p},t)=-\frac{i}{\hbar }b(-\mathbf{p},t)-\frac{i}{2}%
\lambda (\mathbf{p},t)a^{\dag }(\mathbf{p},t)-ev_{F}\sum_{\mathbf{k}%
}\sum_{k_{z}<\varphi k}\frac{i}{\sqrt{2\hbar cVk}}  \nonumber \\
&&\times \left\{ \Gamma _{uv}^{\alpha }(\mathbf{p}+\hbar K,\mathbf{p}%
;t)a^{\dag }(\mathbf{p}+\hbar \mathbf{k},t)+\Gamma _{vv}^{\alpha }(\mathbf{p}%
+\hbar K,\mathbf{p};t)b(-\mathbf{p}^{\prime },t)\right\} \hat{A}^{\alpha }(%
\mathbf{k},k_{z},t),  \nonumber \\
&&\dot{\hat{A}}^{(\pm )\alpha }(\pm \mathbf{k},k_{z},t)=\pm iK\mathcal{%
A^{(\pm )\alpha }}(\pm \mathbf{k},k_{z},t)\mp ev_{F}\sum_{\mathbf{p}}\frac{i%
}{\sqrt{2\hbar cVk}}  \nonumber \\
&&\left\{ \Gamma _{uu}^{\alpha }(\mathbf{p},\mathbf{p}^{\prime };t)a^{\dag }(%
\mathbf{p},t)a(\mathbf{p}^{\prime },t)+\Gamma _{uv}^{\alpha }(\mathbf{p},%
\mathbf{p}^{\prime };t)a^{\dag }(\mathbf{p},t)b^{\dag }(-\mathbf{p}^{\prime
},t)\right.  \nonumber \\
&&+\left. \Gamma _{vu}^{\alpha }(\mathbf{p},\mathbf{p}^{\prime };t)b(-%
\mathbf{p},t)a(\mathbf{p}^{\prime },t)+\Gamma _{vv}^{\alpha }(\mathbf{p},%
\mathbf{p}^{\prime };t)b(-\mathbf{p},t)b^{\dag }(-\mathbf{p}^{\prime
},t)\right\} ,  \label{2.51}
\end{eqnarray}%
where $\mathbf{p}^{\prime }\mathbf{=p-\hbar k}$.

Now we introduce the correlation function $f(\mathbf{p},\mathbf{p}\,^{\prime
};t)$ of the electron-hole subsystem and the one $F_{\alpha \beta }(\mathbf{k%
},\mathbf{k}\,^{\prime };t)$ of the photon subsystem,
\begin{eqnarray}
&&f(\mathbf{p},\mathbf{p}\,^{\prime };t)=\langle 0,\mathrm{in}|a^{\dag }(%
\mathbf{p},t)a(\mathbf{p}\,^{\prime },t)|0,\mathrm{in}\rangle ,  \nonumber \\
&&F_{\alpha \beta }(\mathbf{k},\mathbf{k}\,^{\prime };t)=\sum_{k_{z}<\varphi
k}\langle 0,\mathrm{in}|\hat{A}^{(+)\alpha }(\mathbf{k},k_{z},t)\hat{A}%
^{(-)\beta }(\mathbf{k}\,^{\prime },k_{z}^{\prime },t)|0,\mathrm{in}\rangle .
\label{2.52}
\end{eqnarray}%
In the case of spatially-homogeneous systems the correlation functions,
being written in the coordinate representation, depend on the coordinate
difference only. Their Fourier transforms depend of the corresponding $%
\delta $ functions. Note that under the condition $\varphi \rightarrow 0$\
the momentum distribution of photons does not depend on $k_{z}$\ and the
correlation function $F_{\alpha \beta }(k,k\,^{\prime };t)$ is simply
proportional to the number of state with $k_{z}<\varphi k$, $n\left(
k\right) =\frac{kd}{2\pi }$. Thus, the correlation functions (\ref{2.52})
have the form:%
\begin{eqnarray}
&&f(\mathbf{p},\mathbf{p}^{\prime };t)=f(\mathbf{p},t)\frac{S}{\left( 2\pi
\right) ^{2}}\delta _{\mathbf{p},\mathbf{p}^{\prime }}\ ,  \label{2.53} \\
&&F_{\alpha \beta }(\mathbf{k},\mathbf{k}^{\prime };t)=F(\mathbf{k}%
,t)n\left( k\right) \frac{S}{\left( 2\pi \right) ^{2}}\delta _{\alpha \beta
}\delta _{\mathbf{k},\mathbf{k}^{\prime }\ },  \label{2.54}
\end{eqnarray}%
where%
\begin{equation}
F(\mathbf{k},t)=\frac{1}{2}\sum_{\alpha }F_{\alpha \alpha }(\mathbf{k},%
\mathbf{k}^{\prime };t)|_{\mathbf{k}^{\prime }=\mathbf{k}\ }.  \label{2.55}
\end{equation}%
In Eq. (\ref{2.53}) $f(\mathbf{p},t)$ is \ a superficial density of electron
(hole) numbers with a momentum $\mathbf{p}$ at the time instant $t,$ whereas
$F(\mathbf{k},t)$ in Eq. (\ref{2.54}) is{\large \ }a superficial density of
photons{\large \ }with the wave vector $\mathbf{k}$ at the time instant $t$
per unit state with a given $k_{z}$. In which follows, we use the notation $%
\hat{A}^{\alpha }(k,t)=\hat{A}^{\alpha }(k,0,t)$ for the photon operator
with a given{\large \ }$k_{z}\rightarrow 0$ {\large \ } .

Eqs. (\ref{2.51}) imply the following equations for the introduced
distribution functions:%
\begin{eqnarray}
&&\dot{f}(\mathbf{p},t)=\frac{1}{2}\lambda (\mathbf{p},t)u(\mathbf{p}%
,t)-iev_{F}\sum_{\mathbf{k}}\frac{(2\pi )^{2}n\left( k\right) }{\sqrt{2\hbar
cVk}}  \nonumber \\
&&\times \left\{ \Gamma _{uu}^{\alpha \ast }(\mathbf{p},\mathbf{p}-\hbar
\mathbf{k};t)\langle 0,\mathrm{in}|a^{\dag }(\mathbf{p}-\hbar \mathbf{k},t)a(%
\mathbf{p},t)\hat{A}^{\alpha }(\mathbf{k},t)|0,\mathrm{in}\rangle \right.
\nonumber \\
&&+\Gamma _{uv}^{\alpha \ast }(\mathbf{p},\mathbf{p}-\hbar \mathbf{k}%
;t)\langle 0,\mathrm{in}|b(-\mathbf{p}+\hbar \mathbf{k},t)a(\mathbf{p},t)%
\hat{A}^{\alpha }(\mathbf{k},t)|0,\mathrm{in}\rangle  \nonumber \\
&&-\Gamma _{uu}^{\alpha }(\mathbf{p},\mathbf{p}-\hbar \mathbf{k};t)\langle 0,%
\mathrm{in}|a^{\dag }(\mathbf{p},t)a(\mathbf{p}-\hbar \mathbf{k},t)\hat{A}%
^{\alpha }(\mathbf{k},t)|0,\mathrm{in}\rangle  \nonumber \\
&&-\left. \Gamma _{uv}^{\alpha }(\mathbf{p},\mathbf{p}-\hbar \mathbf{k}%
;t)\langle 0,\mathrm{in}|a^{\dag }(\mathbf{p},t)b^{\dag }(-\mathbf{p}-\hbar
\mathbf{k},t)\hat{A}^{\alpha }(\mathbf{k},t)|0,\mathrm{in}\rangle \right\} ,
\label{2.56}
\end{eqnarray}%
and%
\begin{eqnarray}
&&\dot{F}(\mathbf{k},t)=-iev_{F}\sum_{\mathbf{p}}\frac{2\pi ^{2}}{\sqrt{%
\hbar cVk}}\left\{ \Gamma _{uu}^{\alpha }(\mathbf{p},\mathbf{p}+\hbar
\mathbf{k};t)\langle 0,\mathrm{in}|a^{\dag }(\mathbf{p},t)a(\mathbf{p}+\hbar
\mathbf{k},t)\hat{A}^{(-)\alpha }(\mathbf{k},t)|0,\mathrm{in}\rangle \right.
\nonumber \\
&&+\Gamma _{uv}^{\alpha }(\mathbf{p},\mathbf{p}+\hbar \mathbf{k};t)\langle 0,%
\mathrm{in}|a^{\dag }(\mathbf{p},t)b^{\dag }(-\mathbf{p}-\hbar \mathbf{k},t)%
\hat{A}^{(-)\alpha }(\mathbf{k},t)|0,\mathrm{in}\rangle  \nonumber \\
&&+\Gamma _{vu}^{\alpha }(\mathbf{p},\mathbf{p}+\hbar \mathbf{k};t)\langle 0,%
\mathrm{in}|b(-\mathbf{p},t)a(\mathbf{p}+\hbar \mathbf{k},t)\hat{A}%
^{(-)\alpha }(\mathbf{k},t)|0,\mathrm{in}\rangle  \nonumber \\
&&+\Gamma _{vv}^{\alpha }(\mathbf{p},\mathbf{p}+\hbar \mathbf{k};t)\langle 0,%
\mathrm{in}|b(-\mathbf{p},t)b^{\dag }(-\mathbf{p}-\hbar \mathbf{k},t)\hat{A}%
^{(-)\alpha }(\mathbf{k},t)|0,\mathrm{in}\rangle  \nonumber \\
&&-\Gamma _{uu}^{\alpha }(\mathbf{p},\mathbf{p}-\hbar \mathbf{k};t)\langle 0,%
\mathrm{in}|\hat{A}^{(+)\alpha }(\mathbf{k},t)a^{\dag }(\mathbf{p},t)a(%
\mathbf{p}-\hbar \mathbf{k},t)|0,\mathrm{in}\rangle  \nonumber \\
&&-\Gamma _{uv}^{\alpha }(\mathbf{p},\mathbf{p}-\hbar \mathbf{k};t)\langle 0,%
\mathrm{in}|\hat{A}^{(+)\alpha }(\mathbf{k},t)a^{\dag }(\mathbf{p},t)b^{\dag
}(-\mathbf{p}+\hbar \mathbf{k},t)|0,\mathrm{in}\rangle  \nonumber \\
&&-\Gamma _{vu}^{\alpha }(\mathbf{p,p}-\hbar \mathbf{k};t)\langle 0,\mathrm{%
in}|\hat{A}^{(+)\alpha }(\mathbf{k},t)b(-\mathbf{p},t)a(p-\hbar \mathbf{k}%
,t)|0,\mathrm{in}\rangle  \nonumber \\
&&-\left. \Gamma _{vv}^{\alpha }(\mathbf{p},\mathbf{p}-\hbar \mathbf{k}%
;t)\langle 0,\mathrm{in}|\hat{A}^{(+)\alpha }(\mathbf{k},t)b(-\mathbf{p}%
,t)b^{\dag }(-\mathbf{p}+\hbar \mathbf{k},t)|0,\mathrm{in}\rangle \right\} .
\label{2.57}
\end{eqnarray}

Equation (\ref{2.56}) reduces to the first equation of set (\ref{2.28}) when
the electron-hole system does not interact with photons

The system of Eqs. (\ref{2.56}) and (\ref{2.57}) is not closed, their RHS
contain higher order correlation functions. The means $\langle 0,\mathrm{in}%
|a^{\dag }a\hat{A}^{(\pm )}|0,\mathrm{in}\rangle $ and $\langle 0,\mathrm{in}%
|b^{\dag }b\hat{A}^{(\pm )}|0,\mathrm{in}\rangle $ correspond to processes
of induced irradiation and absorption, whereas the means $\langle 0,\mathrm{%
in}|ab\hat{A}^{(+)}|0,\mathrm{in}\rangle $ and $\langle 0,\mathrm{in}%
|a^{\dag }b^{\dag }\hat{A}^{(-)}|0,\mathrm{in}\rangle $ correspond to
processes with pair annihilation and creation. We do not consider the
phantom process with the correlators $\langle 0,\mathrm{in}|a^{\dag }b^{\dag
}\hat{A}^{(+)}|0,\mathrm{in}\rangle $ and $\langle 0,\mathrm{in}|ab\hat{A}%
^{(-)}|0,\mathrm{in}\rangle $.

Equations (\ref{2.56}) and (\ref{2.57}) constitute a part of the
Bogoliubov--Born--Green--Kirkwood--Yvon (BBGKY) chain of equations in the
electron-hole and photon sectors. To close the set (\ref{2.56}) and (\ref%
{2.57}) it is necessary to obtain equations of the second level for all the
above mentioned correlation functions. These equations will already contain
two-particle correlation functions, which, assuming a weak interaction
between the subsystems, can again be represented through the single-particle
correlator.

Below we give an example of a truncation procedure:
\begin{eqnarray}
&&\ \langle 0,\mathrm{in}|a^{\dag }(\mathbf{p}_{1},t)a(\mathbf{p}_{2},t)\hat{%
A}_{\beta }(\mathbf{k}\,^{\prime },t)\hat{A}_{\alpha }^{(\pm )}(\mathbf{k}%
,t)|0,\mathrm{in}\rangle =\frac{S^{2}}{\left( 2\pi \right) ^{4}}f(\mathbf{p}%
_{1},t)\delta _{\mathbf{p}_{1},\mathbf{p}_{2}}\delta _{\alpha \beta }\left\{
\begin{array}{c}
\left[ F(\mathbf{k},t)+1\right] \delta _{\mathbf{k},-\mathbf{k}^{\prime }}
\\
F(\mathbf{k},t)\delta _{\mathbf{k},\mathbf{k}^{\prime }}%
\end{array}%
\right. ,  \nonumber \\
&&\langle 0,\mathrm{in}|a^{\dag }(\mathbf{p}_{1},t)a(\mathbf{p}%
_{2},t)a^{\dag }(\mathbf{p}\,^{\prime },t)a(\mathbf{p}\,^{\prime \prime
},t)|0,\mathrm{in}\rangle  \nonumber \\
&=&\frac{S^{2}}{\left( 2\pi \right) ^{4}}f(\mathbf{p}_{1},t)\left[ f(\mathbf{%
p}^{\prime },t)\delta _{\mathbf{p}_{1},\mathbf{p}_{2}}\delta _{\mathbf{p}%
^{\prime },\mathbf{p}^{\prime \prime }}+\left[ 1-f(\mathbf{p}^{\prime },t)%
\right] \delta _{\mathbf{p}_{1},\mathbf{p}^{\prime \prime }}\delta _{\mathbf{%
p}_{2},\mathbf{p}^{\prime }}\right] ,  \nonumber \\
&&\ \langle 0,\mathrm{in}|a^{\dag }(\mathbf{p}_{1},t)a(\mathbf{p}_{2},t)b(-%
\mathbf{p}^{\prime },t)b^{\dag }(-\mathbf{p}^{\prime \prime },t)|0,\mathrm{in%
}\rangle =\frac{S^{2}}{\left( 2\pi \right) ^{4}}f(\mathbf{p}_{1},t)\left[
1-f(\mathbf{p}^{\prime },t)\right] \delta _{\mathbf{p}_{1},\mathbf{p}%
_{2}}\delta _{\mathbf{p}^{\prime },\mathbf{p}^{\prime \prime }}\ .
\label{2.58}
\end{eqnarray}%
It corresponds to the random-phase-approximation (RPA) \cite%
{Kadanoff:1962,Zubarev:1996}. We note that in the truncation procedure one
neglects polarization effects in the resulting CI and in the evolution
equations for the polarization functions\textbf{\ $u(\mathbf{p},t)$ }and $%
v(p,t)$ (\ref{2.28}). Here, the assumption of spatial homogeneity and its
consequence (\ref{2.52}) were taken into account.

In such a way, a closed set of equations for the distribution functions $f(%
\mathbf{p},t)$ and $F(\mathbf{k},t)$ can be obtained. In the thermodynamic
limit $V\rightarrow \infty $ it reads:
\begin{eqnarray}
\dot{f}(\mathbf{p},t) &=&I(\mathbf{p},t)+C_{\gamma }(\mathbf{p},t)+C_{%
\mathrm{eh}}(\mathbf{p},t)\ ,  \label{2.59} \\
\dot{F}(\mathbf{k},t) &=&S_{\gamma }(\mathbf{k},t)+S_{\mathrm{eh}}(\mathbf{k}%
,t).  \label{2.60}
\end{eqnarray}%
\

Collision integrals $C_{\gamma }(\mathbf{p},t)$ and $C_{\mathrm{eh}}(\mathbf{%
p},t)$ that describe creation and annihilation of $\mathrm{eh}$-pairs or a
distribution of carries in momenta in course of one-photon absorption or
emission respectively have the form:

\begin{eqnarray}
&&C_{\gamma }(\mathbf{p},t)=2\int \frac{d\mathbf{k}n\left( k\right) }{\left(
2\pi \hbar \right) ^{2}}\int_{t_{0}}^{t}dt^{\prime }K_{\gamma }(\mathbf{p},%
\mathbf{p}+\hbar \mathbf{k};t,t^{\prime })  \nonumber \\
&&\left\{ f(\mathbf{p},t^{\prime })f(\mathbf{p}+\hbar \mathbf{k},t^{\prime
})[1+F(\mathbf{k},t^{\prime })]-[1-f(\mathbf{p},t^{\prime })][1-f(\mathbf{p}%
+\hbar \mathbf{k},t^{\prime })]F(\mathbf{k},t^{\prime })\right\}
\label{2.61} \\
&&C_{\mathrm{eh}}(\mathbf{p},t)=2\int \frac{d\mathbf{k}n\left( k\right) }{%
\left( 2\pi \hbar \right) ^{2}}\int_{t_{0}}^{t}dt^{\prime }K_{\mathrm{eh}}(%
\mathbf{p},\mathbf{p}+\hbar \mathbf{k};t,t^{\prime })  \nonumber \\
&&\left\{ f(\mathbf{p},t^{\prime })[1-f(\mathbf{p}+\hbar \mathbf{k}%
,t^{\prime })][1+F(\mathbf{k},t^{\prime })]-f(\mathbf{p},t^{\prime })[1-f(%
\mathbf{p}+\hbar \mathbf{k},t^{\prime })]F(\mathbf{k},t^{\prime })\right\} .
\label{2.62}
\end{eqnarray}%
Collision integrals in the photon sector read:
\begin{eqnarray}
&&S_{\gamma }(\mathbf{k},t)=2\int \frac{d\mathbf{p}}{\left( 2\pi \hbar
\right) ^{2}}\int_{t_{0}}^{t}dt^{\prime }K_{\gamma }(\mathbf{p},\mathbf{p}%
+\hbar \mathbf{k};t,t^{\prime })  \nonumber \\
&&\left\{ f(\mathbf{p},t^{\prime })f(\mathbf{p}+\hbar \mathbf{k},t^{\prime
})[1+F(\mathbf{k},t^{\prime })]-[1-f(\mathbf{p},t^{\prime })][1-f(\mathbf{p}%
+\hbar \mathbf{k},t^{\prime })]F(\mathbf{k},t^{\prime })\right\}
\label{2.63} \\
&&S_{\mathrm{eh}}(\mathbf{k},t)=2\int \frac{d\mathbf{p}}{\left( 2\pi \hbar
\right) ^{2}}\int_{t_{0}}^{t}dt^{\prime }K_{\mathrm{eh}}(\mathbf{p},\mathbf{p%
}+\hbar \mathbf{k};t,t^{\prime })  \nonumber \\
&&\left\{ f(\mathbf{p},t^{\prime })[1-f(\mathbf{p}+\hbar \mathbf{k}%
,t^{\prime })][1+F(\mathbf{k},t^{\prime })]-f(\mathbf{p},t^{\prime })[1-f(%
\mathbf{p}+\hbar \mathbf{k},t^{\prime })]F(\mathbf{k},t^{\prime })\right\} .
\label{2.64}
\end{eqnarray}

We note that kernels in collision integrals (\ref{2.61})-(\ref{2.64}) in the
photon and $\mathrm{eh}$-sectors are the same and have the form:
\begin{eqnarray}
&&K_{\gamma }(\mathbf{p},\mathbf{p}+\hbar \mathbf{k};t,t^{\prime })=\frac{%
(ev_{F})^{2}}{2\hbar ck}\Gamma _{uv}^{\alpha }(\mathbf{p},\mathbf{p}+\hbar
\mathbf{k};t)\Gamma _{uv}^{\alpha \ast }(\mathbf{p},\mathbf{p}+\hbar \mathbf{%
k};t^{\prime })\cos \Theta ^{(+)}(\mathbf{p},\mathbf{p}+\hbar \mathbf{k}%
;t,t^{\prime }),  \nonumber \\
&&K_{\mathrm{eh}}(\mathbf{p},\mathbf{p}+\hbar \mathbf{k};t,t^{\prime })=%
\frac{(ev_{F})^{2}}{2\hbar ck}\Gamma _{uu}^{\alpha }(\mathbf{p},\mathbf{p}%
+\hbar \mathbf{k};t)\Gamma _{uu}^{\alpha \ast }(\mathbf{p},\mathbf{p}+\hbar
\mathbf{k};t^{\prime })\cos \Theta ^{(-)}(\mathbf{p},\mathbf{p}+\hbar
\mathbf{k};t,t^{\prime }),  \nonumber \\
&&\Theta ^{(\pm )}(\mathbf{p},\mathbf{p}+\hbar \mathbf{k};t,t^{\prime })=%
\frac{1}{\hbar }\int_{t^{\prime }}^{t}d\tau \left[ \varepsilon (\mathbf{p}%
,\tau )\pm \varepsilon (\mathbf{p}+\hbar \mathbf{k},\tau )-c\hbar k\right] .
\label{2.65}
\end{eqnarray}%
This fact allows one to interpret collision integrals (\ref{2.61}), (\ref%
{2.62}) and (\ref{2.63}), (\ref{2.64}) as a reduction of some unified
integrands to the photon and $\mathrm{eh}$-sectors.

\section{Setting of the problem}

\label{III}

The set of KE (\ref{2.59}) and (\ref{2.60}) with CI (\ref{2.61})-(\ref{2.64}%
) and Maxwell equations (\ref{2.37}) describe a self-consistent dynamics of
carriers in the graphene and a behavior of the internal electromagnetic
fields. The same set of equations describes an electromagnetic radiation
from graphene, both classical, generated by plasma currents, and quantum,
due to one-photon processes. Since the classical radiation has already been
studied\footnote{%
We note that the back reaction classical radiation accompanying the particle
production by a slowly varying strong external electric field was evaluated
in Ref. \cite{Gavrilov:2012}. In this case the backreaction field is strong
and slowly varying, as well.} on the basis of KE, see \cite%
{Baudisch:2018,Bowlan:2014}, in the present work we focus our attention on
the quantum radiation, in particular, comparing its characteristics with the
once of the classical radiation. Below, we study this problem solving photon
KE (\ref{2.60}). This corresponds to neglecting the back reaction of the
quantum radiation on carrier dynamics, which is still being described by Eq.
(\ref{2.25a}) but with effective external field (\ref{2.38}). Thus, the
action of the photon subsystem on the evolution of the $\mathrm{eh}$
subsystem and the cascade processes remains outside the scope of this study,
as well as the action of the quantum radiation on effective electric field (%
\ref{2.38}).

In order to solve the problem, we rewrite Eqs. (\ref{2.63}) and (\ref{2.64}%
), neglecting the influence of the photon subsystem,
\begin{eqnarray}
&&S_{\gamma }(\mathbf{k},t)=\frac{2}{\left( 2\pi \hbar \right) ^{2}}\int d%
\mathbf{p}\int_{t_{0}}^{t}dt^{\prime }K_{\gamma }\left( \mathbf{p,p+}\hbar
\mathbf{k};t,t^{\prime }\right) f(\mathbf{p},t^{\prime })f\left( \mathbf{p+}%
\hbar \mathbf{k,}t^{\prime }\right) ,  \label{3.1} \\
&&S_{\mathrm{eh}}(\mathbf{k},t)=\frac{2}{\left( 2\pi \hbar \right) ^{2}}\int
d\mathbf{p}\int_{t_{0}}^{t}dt^{\prime }K_{\mathrm{eh}}\left( \mathbf{p,p+}%
\hbar \mathbf{k};t,t^{\prime }\right) f(\mathbf{p},t^{\prime })\left[ 1-f(%
\mathbf{p+}\hbar \mathbf{k},t^{\prime })\right] .  \label{3.2}
\end{eqnarray}

Thus, at this stage, we study Eq. (\ref{2.60}) with CI (\ref{3.1}) and (\ref%
{3.2}) that do not contain the photon distribution function. We take into
account only a spontaneous emission from the carrier currents produced by an
external electric field. Integrals (\ref{3.1}) and (\ref{3.2}) contain
quadratic combinations of the distribution functions of the carriers. These
functions should be found as a result of solving an auxiliary problem on the
basis of KE (\ref{2.25a}) or equivalent equations (\ref{2.28}).

\section{Studying processes in specific external fields}

\label{IV}

\subsection{Models of external fields}

\label{IVa}

Below, we study the KE that describe the behavior of the carriers and the
electromagnetic field in the graphene when external field is linearly
polarized, $A_{\mathrm{ext}}^{1}=0$ ($E_{\mathrm{ext}}^{1}=0$). Two
following models of the linear polarized short laser pulse \cite%
{Hebenstreit:2009} are convenient for numerical modelling of the CI (\ref%
{3.1}) and (\ref{3.2}). First we consider the two following potentials $A_{%
\mathrm{ext}}^{2}(t)$ and corresponding fields $E_{\mathrm{ext}}^{2}(t)$:%
\begin{eqnarray}
&&A_{\mathrm{ext}}^{2}(t)=A_{c}(t)=-\sqrt{\frac{\pi }{8}}E_{0}\tau \exp
(-\sigma ^{2}/2){\mathrm{erf}}\left( \frac{t}{\sqrt{2\tau }}-i\frac{\sigma }{%
\sqrt{2}}\right) +c.c.\ ,  \nonumber \\
&&E_{\mathrm{ext}}^{2}(t)=E_{c}(t)=E_{0}\exp (-t^{2}/2\tau ^{2})\cos \omega
t,\ \sigma =\omega \tau \ ,\ \sigma =\omega \tau \ ,  \label{4.1}
\end{eqnarray}%
and%
\begin{eqnarray}
A_{\mathrm{ext}}^{2}(t) &=&A_{s}(t)=i\sqrt{\frac{\pi }{8}}E_{0}\tau \exp
(-\sigma ^{2}/2)\mathrm{erf}\left( \frac{t}{\sqrt{2\tau }}-i\frac{\sigma }{%
\sqrt{2}}\right) -c.c.\ ,  \nonumber \\
E_{\mathrm{ext}}^{2}(t) &=&E_{s}(t)=E_{0}\exp (-t^{2}/2\tau ^{2})\sin \omega
t\ ,  \label{4.2}
\end{eqnarray}%
where $\mathrm{erf}\left( x\right) $ is the error function, $\omega =2\pi /T$
is the cyclic frequency and $\tau $ is the envelope pulse length.

We are going also to use a model of a harmonic electric field in some
analytical calculations:
\begin{equation}
A_{\mathrm{ext}}^{2}(t)=A(t)=-(E_{0}/\omega )\sin \omega t,\ E_{\mathrm{ext}%
}^{2}(t)=E(t)=E_{0}\cos \omega t\ .  \label{4.3}
\end{equation}

\subsection{Low-density approximation}

\label{IVb}

Below, to analyze CI (\ref{3.1}) and (\ref{3.2}) by analytical methods we
introduce some approximations for basic constituents of these CI.

In order to estimate of the distribution function, we will use the
low-density approximation $f(\mathbf{p},t)\ll 1$ \cite{Fedotov:2011}, which
immediately allows one to write a solution with the zero initial date $%
f(t_{0})=0$ of Eqs. (\ref{2.25a}) - (\ref{2.27}) as:%
\begin{equation}
f(t)=\frac{1}{2}\int_{t_{0}}^{t}dt^{\prime }\lambda (t^{\prime
})\int_{t_{0}}^{t^{\prime }}dt^{\prime \prime }\lambda (t^{\prime \prime
})\cos \theta (t^{\prime },t^{\prime \prime }).  \label{4.4}
\end{equation}%
It is convenient to rewrite this solution in a slightly different form using
definition (\ref{2.27}) of the phase and setting the initial time moment
(the time of the switching on the external field) to minus infinity ($%
t_{0}=-\infty )$,%
\begin{eqnarray}
&&f(t)=\frac{1}{2}\int_{-\infty }^{t}dt^{\prime }\lambda _{c}(t^{\prime
})\int_{-\infty }^{t^{\prime }}dt^{\prime \prime }\lambda _{c}(t^{\prime
\prime })+\frac{1}{2}\int_{-\infty }^{t}dt^{\prime }\lambda _{s}(t^{\prime
})\int_{-\infty }^{t^{\prime }}dt^{\prime \prime }\lambda _{s}(t^{\prime
\prime })\ ,  \nonumber \\
&&\lambda _{c}(t)=\lambda (t)\cos \theta (t,-\infty ),\ \ \lambda
_{s}(t)=\lambda (t)\sin \theta (t,-\infty ).  \label{4.5}
\end{eqnarray}%
Then solution (\ref{4.5}) can be represented as:%
\begin{equation}
f(t)=\frac{1}{4}\left[ \int_{-\infty }^{t}dt^{\prime }\lambda _{s}(t^{\prime
})\right] ^{2}+\frac{1}{4}\left[ \int_{-\infty }^{t}dt^{\prime }\lambda
_{c}(t^{\prime })\right] ^{2},  \label{4.6}
\end{equation}%
which immediately implies that $f(t)\geq 0$.

Then we introduce the approximation of the effective electromagnetic mass
using more accurate estimates of the regularized energy \cite%
{Mostepanenko:1974}:
\begin{eqnarray}
&&\varepsilon (\mathbf{p},t)\rightarrow \varepsilon _{\ast }(\mathbf{p}%
)=\varepsilon _{\ast }(p)=v_{F}\sqrt{m_{\ast }^{2}v_{F}^{2}+p^{2}},
\nonumber \\
&&m_{\ast }^{2}=\frac{e^{2}}{c^{2}v_{F}^{2}}\frac{1}{2T}%
\int_{-T}^{T}dtA^{2}(t),\ p=|\mathbf{p}|\ ,  \label{4.7}
\end{eqnarray}%
where $T=2\pi /\omega $ is the period of field oscillations with cyclic
frequency $\omega =2\pi \nu $. In the periodic field model (\ref{4.3}), this
mass reads:%
\begin{equation}
m_{\ast }=\frac{eE_{0}}{\sqrt{2}v_{F}\omega }.  \label{4.8}
\end{equation}%
One can present estimates of $m_{\ast }$ using parameters that appeared in
two experimental studies: $m_{\ast }=0.03\ m_{e}$ for $E_{0}=3\cdot 10^{6}\
V/m$,\ $\nu =2\cdot 10^{12}\ Hz$ \cite{Bowlan:2014};\ $m_{\ast }=0.047\
m_{e},\ E_{0}=2.3\cdot 10^{8}\ V/m$,\ $\nu =96.7\cdot 10^{12}\ Hz$ \cite%
{Baudisch:2018}.

In the above mentioned approximation, the phase $\theta (t,t^{\prime })$ and
amplitude $\lambda (\mathbf{p},t)$ (\ref{2.27}) are:%
\begin{equation}
\theta (t,t_{0})=\frac{2\varepsilon _{\ast }}{\hbar }(t-t_{0}),\ \lambda
_{\ast }(\mathbf{p},t)=eE(t)l_{\ast },\ l_{\ast }=-v_{F}^{2}p_{1}\varepsilon
_{\ast }^{-2}\ .  \label{4.9}
\end{equation}%
This leads to the fact that only two main harmonics remain in amplitudes (%
\ref{4.5}), $\Omega _{\pm }=2\varepsilon _{\ast }/\hbar \pm \omega $, such
that:
\begin{equation}
\lambda _{\ast c}(t)=\frac{1}{2}eE_{0}l_{\ast }\left[ \cos \Omega _{+}t+\cos
\Omega _{-}t\right] ,\ \lambda _{\ast s}(t)=\frac{1}{2}eE_{0}l_{\ast }\left[
\sin \Omega _{+}t+\sin \Omega _{-}t\right] \ .  \label{4.10}
\end{equation}%
Substituting Eqs. (\ref{4.10}) into Eq. (\ref{4.6}), we obtain the
distribution function for lengthy impulse $\tau \gg T$:%
\begin{equation}
f(\mathbf{p},t)=\left( \frac{eE_{0}l_{\ast }}{4\Omega _{+}\Omega _{-}}%
\right) ^{2}(\Omega _{+}^{2}+\Omega _{-}^{2}+2\Omega _{+}\Omega _{-}\cos
\omega t).  \label{4.11}
\end{equation}%
Taking into account the definition of $\Omega _{\pm },$ we arrive to the
representation:
\begin{equation}
f(\mathbf{p},t)=f^{(0)}(\mathbf{p})+f^{(2)}(\mathbf{p},t),  \label{4.12}
\end{equation}%
where the function
\begin{equation}
f^{(0)}(\mathbf{p})=\frac{(e\hbar E_{0}l_{\ast })^{2}(4\varepsilon _{\ast
}^{2}+\hbar ^{2}\omega ^{2})}{8(4\varepsilon _{\ast }^{2}-\hbar ^{2}\omega
^{2})^{2}}  \label{4.13}
\end{equation}%
corresponds to a stationary background distribution and the function
\begin{equation}
f^{(2)}(\mathbf{p},t)=\frac{(e\hbar E_{0}l_{\ast })^{2}}{8(4\varepsilon
_{\ast }^{2}-\hbar ^{2}\omega ^{2})}\cos 2\omega t  \label{4.14}
\end{equation}%
correspond to the breathing mode on the doubled frequency{\Huge \ }of the
external field.

In the general case, the distribution function $f(\mathbf{p},t)$ contains
the even harmonies of external field only. This conclusion follows from the
general structure of the basic KE (\ref{2.25a}), (\ref{2.27}) or its
alternative form (\ref{2.28}).

\subsection{Calculating kernels (\protect\ref{2.65})}

\label{IVc}

Let us estimate at first convolutions of the matrix vertices of functions of
type (\ref{2.50}) entering in kernels (\ref{2.65}). To this end, we
calculate the functions themselves componentwise, using definitions of
evolution operator (\ref{2.7}) and spinors (\ref{2.8}),
\begin{eqnarray}
\Gamma _{uu}^{1}(\mathbf{p},\mathbf{p}_{1};t) &=&\cos [\varkappa (\mathbf{p}%
)/2+\varkappa (\mathbf{p}_{1})/2]=-i\Gamma _{uv}^{2}(\mathbf{p},\mathbf{p}%
_{1};t),  \nonumber \\
\Gamma _{uu}^{2}(\mathbf{p},\mathbf{p}_{1};t) &=&\sin [\varkappa (\mathbf{p}%
)/2+\varkappa (\mathbf{p}_{1})/2]=i\Gamma _{uv}^{1}(\mathbf{p},\mathbf{p}%
_{1};t),  \label{4.15}
\end{eqnarray}%
where $\varkappa (\mathbf{p})=\arctan (P^{2}/P^{1})$. It is convenient to
write these functions as some algebraic expressions of the quasimomenta$\
P^{1}$ and $P^{2}.$ It should be noted that functions (\ref{4.15}) describe
the influence of the external field on elementary acts of interaction of
electrons and holes (considered as quasiparticles) with photons. Each of the
functions is periodic with the period of the external field, and their
convolutions in kernels of CI (\ref{2.63}), (\ref{2.64}) depend on the
observation time $t$ and the antecedent time $t^{\prime }$ that describes
memory effects in the interaction. Sum rules at coinciding times $%
t=t^{\prime }$ follow from Eqs. (\ref{4.15}):%
\begin{equation}
\sum_{\alpha }|\Gamma _{uu}^{\alpha }(\mathbf{p},\mathbf{p}%
_{1};t)|^{2}=\sum_{\alpha }|\Gamma _{vv}^{\alpha }(\mathbf{p},\mathbf{p}%
_{1};t)|^{2}=\sum_{\alpha }|\Gamma _{uv}^{\alpha }(\mathbf{p},\mathbf{p}%
_{1};t)|^{2}=1.  \label{4.16}
\end{equation}%
Here, the influence of the external field is taken into account, but, in
comparison with convolutions in kernels (\ref{2.65}), the retardation effect
is neglected. Such an approximation corresponds to the neglecting harmonics
of the external field. The intensity of these harmonics at the fundamental
frequency is small in comparison with (\ref{4.16}),%
\begin{equation}
\frac{ev_{F}^{2}p_{\parallel }E_{0}}{c\omega \varepsilon _{\ast }^{2}(%
\mathbf{p})}\ll 1,  \label{4.17}
\end{equation}%
where $p_{\parallel }=\mathbf{p}\mathbf{k}/k$ is the longitudinal momentum.
Properties (\ref{4.16}) will be used below for estimates of kernels (\ref%
{2.65}) of the CI.

Electromagnetic mass approximation (\ref{4.7}) is also effective in
estimating phases in kernels of CI (\ref{2.65}). In this approximation%
\begin{equation}
\Theta _{\ast }^{(\pm )}(\mathbf{p},\mathbf{p}_{1};t,t^{\prime })=\left[
\varepsilon _{\ast }(\mathbf{p})\pm \varepsilon (\mathbf{p}_{1})-c\hbar k%
\right] (t-t^{\prime }).  \label{4.18}
\end{equation}

Usually, in the absence of external fields, deriving KE (see e.g. \cite%
{deGroot:1980,Kadanoff:1962,Zubarev:1996}), after integration over the time
in CI, one obtains energy conservation laws in elementary acts of scattering
of constituents. In the considered highly nonequilibrium situation, phase (%
\ref{4.18}) can be modified by harmonics of the external field due to the
time dependence of the carrier distribution function (an example is function
(\ref{4.12}). It can influence significantly upon elementary processes in
Eq. (\ref{4.18}). For example, the spontaneous single-photon annihilation
channel (sign "$+$" in Eq. (\ref{4.18})) is forbidden at $m_{\ast }\neq 0$
but the presence of an external field may lead to its opening (process of
the stimulated annihilation). Such situation is known also in the standard
strong field QED \cite{Amirov:1974,Seminozhenko:1982}. It corresponds to a
general theory of an external field influence on scattering process in
strong nonequilibrium systems, see \cite%
{Zubarev:1996,Amirov:1974,Seminozhenko:1982,Smolyansky:2020,Smolyansky:2020a}%
.

Now we can advance in calculating time integrals in CI in the leading
harmonic approximation under consideration. Taking into account the
structure of CI as functionals of the distribution function $f(\mathbf{p},t)$
(\ref{4.12}) - (\ref{4.14}), phase (\ref{4.18}) can acquire additional
contributions $n\omega $,$\ n=0,\pm 2,\pm 4$ from higher harmonics of the
external field. Then integration in time leads to the appearance of singular
functions determined by roots of the equation:%
\begin{equation}
\varphi ^{(\pm )}(\mathbf{p},\mathbf{k};n)=\varepsilon _{\ast }(\mathbf{p}%
)\pm \varepsilon _{\ast }(\mathbf{p}+\hbar \mathbf{k})-c\hbar k+n\hbar
\omega =0.  \label{4.19}
\end{equation}%
Bellow these equations are analyzed in some particular cases.

\section{Spectral composition of quantum radiation}

\label{V}

According to KE (\ref{2.60}) the quantum radiation is formed in the
annihilation channel with CI (\ref{2.61}) and in the channel of momentum
redistribution with CI (\ref{2.62}). We consider these channels using
approximate CI (\ref{3.1}) and (\ref{3.2}) in the case of a linearly
polarized electric field, neglecting the retardation in the convolutions of
the vertex functions with substitutions of the exact convolutions by
single-time expressions (\ref{4.16}), taking also into account the
additional approximations discussed earlier in Sect. \ref{IV}:

\begin{itemize}
\item 

\item the low-density approximation for estimation of carrier distribution
functions (\ref{4.12}) - (\ref{4.14});

\item approximation of effective electromagnetic mass (\ref{4.7}) in the
estimations of the distribution function and phases (\ref{2.65}).
\end{itemize}

We note that both CI (\ref{3.1}) and (\ref{3.2}) are functionals of
different degrees of nonlinearity under distribution function. They have to
be calculated in one way in the biharmonic approximation relatively to the
external field frequency.

\subsection{Annihilation channel}

\label{Va}

Let us chose time independent and biharmonic parts of CI (\ref{3.1}),
\begin{eqnarray}
&&S_{\gamma }(\mathbf{k},t)=S_{\gamma }^{(0)}(\mathbf{k})+S_{\gamma }^{(2)}(%
\mathbf{k},t),  \label{5.1} \\
&&S_{\gamma }^{(0)}(\mathbf{k})=\frac{(ev_{F})^{2}}{2\hbar cka}\int \frac{d%
\mathbf{p}}{(2\pi \hbar )^{2}}\int^{t}dt^{\prime }\cos \Theta _{\ast
}^{\left( +\right) }(\mathbf{p},\mathbf{p}+\hbar \mathbf{k};t,t^{\prime
})f^{\left( 0\right) }(\mathbf{p})f^{(0)}(\mathbf{p}+\hbar \mathbf{k}),
\label{5.2} \\
&&S_{\gamma }^{(2)}(\mathbf{k},t)=\frac{(ev_{F})^{2}}{2\hbar cka}\int \frac{d%
\mathbf{p}}{(2\pi \hbar )^{2}}\int^{t}dt^{\prime }\cos \Theta _{\ast
}^{\left( +\right) }(\mathbf{p},\mathbf{p}+\hbar \mathbf{k};t,t^{\prime })
\nonumber \\
&&\times \left[ f^{(0)}(\mathbf{p})f^{(2)}(\mathbf{p}+\hbar \mathbf{k}%
)+f^{(2)}(\mathbf{p})f^{(0)}(\mathbf{p}+\hbar \mathbf{k})\right] \cos
2\omega t^{\prime }.  \label{5.3}
\end{eqnarray}%
Here the functions $f^{(0)}$ and $f^{(2)}$ are given by Eqs. (\ref{4.12}) - (%
\ref{4.14}).

The time integral in CI (\ref{5.2}) with phase (\ref{4.18}) (upper sign "$+$%
") implies the following time independent result:
\begin{equation}
\int_{-\infty }^{t}dt^{\prime }\cos \Theta _{\ast }^{\left( +\right) }(%
\mathbf{p},\mathbf{p}+\hbar \mathbf{k};t,t^{\prime })=\pi \hbar \delta \left[
\varepsilon _{\ast }(\mathbf{p})+\varepsilon _{\ast }(\mathbf{p}+\hbar
\mathbf{k})-\hbar ck\right] .  \label{5.4}
\end{equation}%
Here the $\delta $ - function reflects the energy conservation law of the
single - photon annihilation process which cannot be realized at $m_{\ast
}\neq 0$ (see Sect. \ref{IV}). In other words Eq. (\ref{4.19}) $\varphi
^{(\pm )}(\mathbf{p},\mathbf{k};n=0)=0$ has no physical roots and therefore:
\begin{equation}
S_{\gamma }^{(0)}(\mathbf{k})=0.  \label{5.5}
\end{equation}

The breathing mode is described by CI (\ref{5.3}). Integration here over the
time leads to two types of singularities:
\begin{eqnarray}
&&\int_{-\infty }^{t}dt^{\prime }\cos \Theta _{\ast }^{\left( +\right) }(%
\mathbf{p},\mathbf{p}+\hbar \mathbf{k};t,t^{\prime })\cos 2\omega t^{\prime
}=\pi \hbar \left\{ \delta \left[ \varphi ^{(+)}(\mathbf{p},\mathbf{k};n=2)%
\right] \right.  \label{5.6} \\
&&\left. +\delta \left[ \varphi ^{(+)}(\mathbf{p},\mathbf{k};n=-2)\right]
\right\} \cos 2\omega t+\hbar \mathcal{P}\left[ \frac{1}{\varphi ^{(+)}(%
\mathbf{p},\mathbf{k};n=2)}-\frac{1}{\varphi ^{(+)}(\mathbf{p},\mathbf{k}%
;n=-2)}\right] \sin 2\omega t,  \nonumber
\end{eqnarray}%
where $\mathcal{P}$ is the symbol of principal value. The equations
\begin{equation}
\varphi ^{(+)}(\mathbf{p},\mathbf{k};n=\pm 2)=0  \label{5.7}
\end{equation}%
describe two singular energy surfaces in the momentum space which correspond
to the two annihilation processes with participation of higher harmonics of
the external field, namely, an emission of an annihilation photon is
accompanied by simultaneously radiation $(n=-2)$ or absorption $(n=2)$ of
harmonics from reservoir of the external field. One can speak about emission
of "soft" or "hard" photons of the quantized field.

The soft photon annihilation process in integral (\ref{5.6}) is described by
energy condition (\ref{5.7}) with $n=-2$. The long-wave approximation is
warranted in this region of wave numbers:
\begin{equation}
\varepsilon _{\ast }(\mathbf{p}+\hbar \mathbf{k})\simeq \varepsilon _{\ast }(%
\mathbf{p})+\hbar \mathbf{k}\ \partial \varepsilon _{\ast }(\mathbf{p}%
)/\partial \mathbf{p}.  \label{5.8}
\end{equation}%
Together with the condition $v_{F}\ll c$ it leads to the solutions $%
p^{(1,2)}=\pm p_{\ast }$ of Eqs. (\ref{5.7}) with $n=-2$, where
\begin{equation}
p_{\ast }=\hbar (ck+2\omega )/2v_{F}.  \label{5.9}
\end{equation}%
Then using the representation
\begin{equation}
\delta \left[ \varphi (p)\right] =\sum_{i}\delta (p-p_{i})\left( \frac{%
d\varphi (p)}{dp}\right) _{p=p_{i}},  \label{5.10}
\end{equation}%
one can specify the $\delta $-function $\delta \left[ \varphi ^{(+)}(\mathbf{%
p},\mathbf{k};n=-2)\right] $ in integral (\ref{5.6}) using the additional
approximation $k\ll 2\omega /c$, to obtain:
\begin{equation}
S_{\gamma }^{(2)(-)^{\prime }}(\mathbf{k},t)=\frac{\pi ^{3}\alpha
^{3}c^{2}v_{F}^{4}E_{0}^{4}}{16\hbar ^{2}ka\omega ^{7}}\left[ \frac{3}{4}%
+\left( \frac{v_{F}k}{\omega }\right) ^{2}\cos ^{2}\varphi \right] \cos
2\omega t,  \label{5.11}
\end{equation}%
where $\varphi $ is the angle between the vectors $\mathbf{k}$ and $\mathbf{E%
}(t)$ characterizing the direction of the radiation.

In the same approximation one can calculate the principal value integral
corresponding to the function $\varphi ^{(+)}(\mathbf{p},\mathbf{k};n=-2)$
in Eq. (\ref{5.6}),
\begin{equation}
S_{\gamma }^{(2)(-)^{\prime \prime }}(\mathbf{k},t)=-\frac{5\alpha
^{3}c^{2}v_{F}^{4}E_{0}^{4}}{32\hbar ^{2}ka\omega ^{7}}\ln \frac{2\hbar
\omega }{m_{\ast }v_{F}^{2}}\cdot \sin 2\omega t.  \label{5.12}
\end{equation}

Thus, the photon production rate in the annihilation channel in the
long-wave approximation is equal to the sum of CI (\ref{5.11}) and (\ref%
{5.12}),
\begin{equation}
S_{\gamma }^{(2)}(\mathbf{k},t)=S_{\gamma }^{(2)(-)^{\prime }}(\mathbf{k}%
,t)+S_{\gamma }^{(2)(-)^{\prime \prime }}(\mathbf{k},t).  \label{5.13}
\end{equation}

\subsection{Channel of the momentum redistribution}

\label{Vb}

The channel of the momentum redistribution is described by CI (\ref{3.2}).
In the low-density limit $f\ll 1$ this CI is a linear functional with
respect to the distribution function $f(\mathbf{p},t)$ and can be
represented by a decomposition of type (\ref{5.1}). Its stationary
background part vanishes by virtue of a violation of the energy conservation
law $\varphi ^{(-)}(\mathbf{p},\mathbf{k};n=0)=0$.

Let us consider the breathing mode of CI (\ref{3.2}). In the approximation
under consideration it reads:
\begin{equation}
S_{\mathrm{eh}}^{(2)}(\mathbf{k},t)=\frac{(ev_{F})^{2}}{(2\pi )^{2}\hbar
^{3}cka}\int d\mathbf{p}\int_{t_{0}\rightarrow -\infty }^{t}dt^{\prime
}f^{(2)}(\mathbf{p},t^{\prime })\cos \Theta _{\ast }^{\left( -\right) }(%
\mathbf{p},\mathbf{p}+\hbar \mathbf{k};t,t^{\prime }),  \label{5.14}
\end{equation}%
where the phase $\Theta _{\ast }^{(-)}$ is defined by Eq. (\ref{4.18}).
Integration over the time gives the following result:
\begin{eqnarray}
&&\int_{-\infty }^{t}dt^{\prime }\cos \Theta _{\ast }^{\left( -\right) }(%
\mathbf{p},\mathbf{p}+\hbar \mathbf{k};t,t^{\prime })\cos 2\omega t^{\prime
}=\frac{\pi \hbar }{2}\left\{ \delta \left[ \varphi ^{(-)}(\mathbf{p},%
\mathbf{k};n=2)\right] \right. +  \label{5.15} \\
&&\left. +\delta \left[ \varphi ^{(-)}(\mathbf{p},\mathbf{k};n=-2)\right]
\right\} \cos 2\omega t+\frac{\hbar }{2}\mathcal{P}\left[ \frac{1}{\varphi
^{(-)}(\mathbf{p},\mathbf{k};n=2)}-\frac{1}{\varphi ^{(-)}(\mathbf{p},%
\mathbf{k};n=-2)}\right] \sin 2\omega t.  \nonumber
\end{eqnarray}

Two equations
\begin{equation}
\varphi ^{(-)}(\mathbf{p},\mathbf{k};n=\pm 2)=0  \label{5.16}
\end{equation}%
define the singular energy surfaces in integral (\ref{5.15}).

The case $n=-2$ corresponds to synchronous emission by a carrier of one
radiation photon with energy $\hbar ck$ and the harmonic with doubled
frequency of the external field. Such process is forbidden.

The case $n=2$ (emission of a photon with capture of the harmonic from the
external field reservoir) is open and is considered bellow in long-wave
approximation (\ref{5.8}). Eq. (\ref{5.16}) for $n=2$ implies the following
condition:
\begin{equation}
\pm \hbar kp_{\parallel }\frac{v_{F}^{2}}{\varepsilon _{\ast }(\mathbf{p})}%
+2\hbar \omega =\hbar ck,  \label{5.17}
\end{equation}%
where $p_{\parallel }$ is the longitudinal momentum.{\Huge \ }It is written
in the form where the LHS represents a change of the carrier energy in
course of the emission of the radiation photon. In order to analyze relation
(\ref{5.17}), we note that it breaks down strongly at $\omega =0$,

\begin{equation}
\hbar kp_{\parallel }\frac{v_{F}^{2}}{\varepsilon _{\ast }(\mathbf{p})}\ll
\hbar ck  \label{5.18}
\end{equation}%
such that a rather high frequency is necessary for the restoration of
equality (\ref{5.17}). Because the LHS of inequality (\ref{5.18}) is very
small in comparison with RHS, one can conclude that the energy of the
radiated photon and the energy $2\hbar \omega $ from the reservoir of an
external field are correlated so that $ck\geq 2\omega $. It means that the
radiated photon captures most of the energy from the external
electromagnetic reservoir.

Now we can introduce the detuning
\begin{equation}
\Delta =\frac{c}{v_{F}}\left( 1-\frac{2\omega }{ck}\right) \geq 0
\label{5.19}
\end{equation}%
and to write a solution of Eq. (\ref{5.17}) in the linear approximation
relatively to the longitudinal momentum $p_{\parallel }$ in the following
form:
\begin{equation}
p_{\parallel }=\pm p_{\parallel }^{\ast },\quad p_{\parallel }^{\ast }=\frac{%
\varepsilon _{\ast \perp }(p_{\perp })}{v_{F}}\Delta ,  \label{5.20}
\end{equation}%
where $\varepsilon _{\ast \perp }(p_{\perp })=v_{F}\sqrt{m_{\ast
}^{2}v_{F}^{2}+p_{\perp }^{2}}$ is the transversal energy.

Let us rewrite CI (\ref{5.14}) keeping in Eq. (\ref{5.15}) the contribution
with the $\delta \left[ \varphi ^{(-)}(\mathbf{p},\mathbf{k};n=2)\right] $
only,
\begin{equation}
S_{\mathrm{eh}}^{(2)^{\prime }}(\mathbf{k},t)=\frac{e^{4}E_{0}^{2}v_{F}^{6}}{%
64\pi cka}\int d^{2}p\frac{p_{1}^{2}}{\varepsilon _{\ast }^{4}(\mathbf{p})}%
\frac{\delta \left[ \varphi ^{(-)}(\mathbf{p},\mathbf{k};n=2)\right] }{%
4\varepsilon _{\ast }^{2}(\mathbf{p})-\hbar ^{2}\omega ^{2}}\cos 2\omega t.
\label{5.21}
\end{equation}%
Here the relation
\begin{equation}
\delta \left[ \varphi ^{(-)}(\mathbf{p},\mathbf{k};n=2)\right] =\frac{%
\varepsilon _{\ast \perp }(p_{\perp })}{kv_{F}^{2}}\left[ \delta
(p_{\parallel }-p_{\parallel }^{\ast })+\delta (p_{\parallel }+p_{\parallel
}^{\ast })\right]  \label{5.22}
\end{equation}%
which is valid in long-wave approximation (\ref{5.8}) was taken into account
. After integration over $p_{\parallel }$, it is conveniently to write the
residual integral using the dimensionless variable $x=\varepsilon _{\ast
\perp }(p_{\perp })/m_{\ast }v_{F}^{2}\ (x\gg 1)$,%
\begin{eqnarray}
&&S_{\mathrm{eh}}^{(2)^{\prime }}(\mathbf{k},t)=\frac{\pi c\hbar \alpha
^{2}E_{0}^{2}}{16k^{2}a[1+\Delta ^{2}]^{5/2}m_{\ast }^{2}v_{F}^{3}}%
\int_{1}^{\infty }\frac{dx}{x^{2}(x^{2}-1)^{1/2}}  \label{5.23} \\
&&\ \times \frac{\Delta ^{2}x^{2}\cos ^{2}\varphi +(x^{2}-1)\sin ^{2}\varphi
}{x^{2}-\xi ^{2}}\cos 2\omega t,\ \xi =\frac{\hbar \omega }{2m_{\ast
}v_{F}^{2}\sqrt{1+\Delta ^{2}}}.  \nonumber
\end{eqnarray}

If we suppose in addition that the inequality $2\omega >ck$ (which that is
equivalent to $\left\vert \Delta \right\vert \gg 1)$ holds true, CI (\ref%
{5.23}) will be equal:
\begin{eqnarray}
&&S_{\mathrm{eh}}^{(2)^{\prime }}(\mathbf{k},t)=\frac{\pi c\hbar }{%
16ka\omega }\left( \frac{\alpha E_{0}}{m_{\ast }v_{F}}\right) ^{2}I(\tilde{%
\xi})\cos ^{2}\varphi \cdot \cos 2\omega t,  \label{5.25} \\
&&I(\tilde{\xi})=\int_{1}^{\infty }\frac{dx}{\sqrt{x^{2}-1}}\frac{1}{x^{2}-%
\tilde{\xi}^{2}}=\frac{\arcsin (\tilde{\xi})}{\tilde{\xi}\sqrt{1-\tilde{\xi}%
^{2}}},\ \tilde{\xi}=\hbar k/4m_{\ast }v_{F}.  \label{5.26}
\end{eqnarray}

The residual part of CI (\ref{5.14}) is a principal-value integral
corresponding to the contribution $\mathcal{P}\left[ \varphi ^{(-)}(\mathbf{p%
},\mathbf{k};n=2)\right] ^{-1}$ in Eq. (\ref{5.15}),
\begin{equation}
S_{\mathrm{eh}}^{(2)^{\prime \prime }}(\mathbf{k},t)=\frac{\alpha ^{2}\hbar
cE_{0}^{2}v_{F}^{4}}{4k^{2}a}\int \frac{d^{2}p}{\varepsilon _{\ast }^{3}(%
\mathbf{p})}\frac{p_{1}^{2}}{4\varepsilon _{\ast }^{2}(\mathbf{p})-\hbar
^{2}\omega ^{2}}\mathcal{P}\left[ \frac{1}{p_{\parallel }-p_{\parallel
}^{\ast }}+\frac{1}{p_{\parallel }+p_{\parallel }^{\ast }}\right] \cos
2\omega t.  \label{5.27}
\end{equation}%
Here $p_{1}=p_{\parallel }\cos \varphi +p_{\perp }\sin \varphi $, where $%
p_{\parallel }$ and $p_{\perp }$ are the longitudinal and transversal
components. In deriving Eq. (\ref{5.27}), it was taken into account
splitting (\ref{5.17}) of the energy surface $\varphi ^{(-)}(\mathbf{p},%
\mathbf{k};n=2)=0$. In this form, it is easy to see that CI (\ref{5.27}) is
equal to zero by virtue of the oddness of the integrand function relatively
substitutions $p_{\parallel }\rightarrow -p_{\parallel }$ and $p_{\perp
}\rightarrow -p_{\perp }$, such that:
\begin{equation}
S_{\mathrm{eh}}^{(2)^{\prime \prime }}(\mathbf{k},t)=0.  \label{5.28}
\end{equation}

\subsection{Comments on the results obtained}

\label{Vc}

Representations of CI (\ref{5.11}) - (\ref{5.13}) and (\ref{5.23}) allow us
to make the following comments and conclusions:

(1) The channel of radiation based on the mechanism of the momentum
redistribution is more effective in comparison with the annihilation
channel. In order to explain this conclusion, it is necessary to note, that
properties of the \textrm{eh}-plasma are defined by the distribution
function $f(\mathbf{p},t)$. In our study this function is calculated in the
low-density approximation, where according to Eqs. (\ref{4.12}) - (\ref{4.14}%
), $f(\mathbf{p},t)\sim \alpha {E_{0}}^{2}$. One can see that CI (\ref{3.1})
in the annihilation channel is a quadratic functional with respect to $f(%
\mathbf{p},t)$ while CI (\ref{3.2}) in the momentum redistribution channel
is the linear functional (in the low-density approximation) with respect to $%
f(\mathbf{p},t)$. It is stipulated by distinctions in these mechanisms of
radiation: emission of photons in the case of the momentum redistribution
takes place with participation of the quasiparticles from one of the
subsystems (subsystem of electrons or holes), whereas radiation in the
annihilation channel is in need in two partners from different subsystems.
Thus, in the leading approximation we have: $S_{\gamma }/S_{\mathrm{eh}}\sim
\alpha $.

(2) A time independent component is absent in the spectrum of the radiation
of the quantized electromagnetic field. In the approximation under
consideration, it is stipulated by the absence of the energy feeding of the
corresponding single-photon processes from the photon reservoir of the
external field.

The basic breathing harmonic of the radiation is the doubled harmonic of the
external field. It is basic frequency of oscillations of the \textrm{eh}%
-plasma (see Eq.(\ref{4.14})). The presence of odd harmonics in the spectrum
of quantized radiation has a principal importance for an experimental
identification of the quantum radiation since the competitive quasiclassical
radiation on the frequency of the plasma oscillations contain odd harmonics
only \cite{Baudisch:2018,Bowlan:2014,Smolyansky:2020b}.

(3) Representations for CI (\ref{5.11}) - (\ref{5.13}) and (\ref{5.23}) were
obtained in the long-wave approximation and describe the situation well in
the area of small wave numbers. At the same time, CI (\ref{5.11}) and (\ref%
{5.12}) of the annihilation channel demonstrate $1/k$ dependence in an
explicit form with feebly marked anisotropy. In the standard QED such $1/k$
behavior was predicted long ago in Refs. \cite{Blaschke:2011,Otto:2017}. The
$k$-dependence in the momentum redistribution channel is more complicated,
it corresponds to the emission of electrons and holes which are accelerated
by the electric field in the opposite directions (see Eq.(\ref{5.20})).

(4) The quantized electromagnetic radiation expands in the graphene plane
and is anisotropic: strong anisotropic effect in the momentum redistribution
channel overlaps on isotropic radiation in the annihilation channel. We note
that the quasiclassical radiation of the plasma oscillations in the graphene%
\textrm{\ }propagates always in the perpendicular direction to the graphene
plane.

\begin{figure}[tbp]
\centering
\includegraphics[width=13 cm]{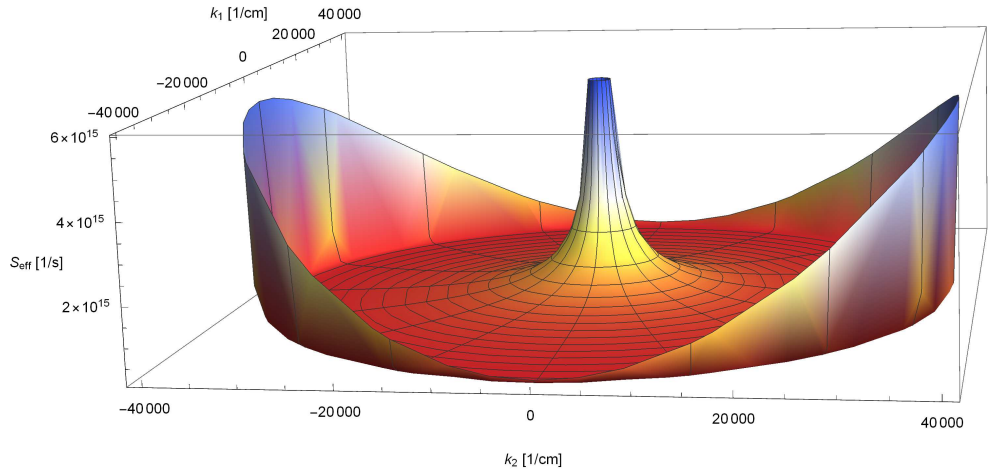}
\caption{The spectral composition of the total photon production rate $S_{%
\protect\gamma} (\mathbf{k}) + S_{\mathrm{eh }} (\mathbf{k})$ in the
long-wave region: the central $1/k$ peak corresponds to the annihilation
channel, the peripheral distribution represents the momentum redistribution
mechanism. }
\label{F1}
\end{figure}

(5) Fig. \ref{F1} illustrates some of these features of the quantized field
radiation. Here the spectral composition of the total photon production rate
in the long waves region is represented. The central isotropic peak
corresponds to the $1/k$-dependence. This feature in the radiation spectrum
corresponds to the annihilation mechanism.

The peripheral strong anisotropic distribution is a result of a momentum
redistribution mechanism in CI (\ref{5.23}) for
$1.95\cdot 10^{-4}\leq \tilde{\xi}\leq 0.456$.{\Huge \ } Such fixation of
the parameter guarantees to find the parameters $E_{0},\ k$ and $\omega $
limited by the usability condition (\ref{5.19}) in the physical region and
provides a sufficient distance from the resonance point $\tilde{\xi}=1$
(see, e.g., Eq. (\ref{5.26})). Fig.\ref{F1} shows that photons are created
predominantly in the direction of the acting external field.

\begin{figure}[H]
\centering
\includegraphics[width=13 cm]{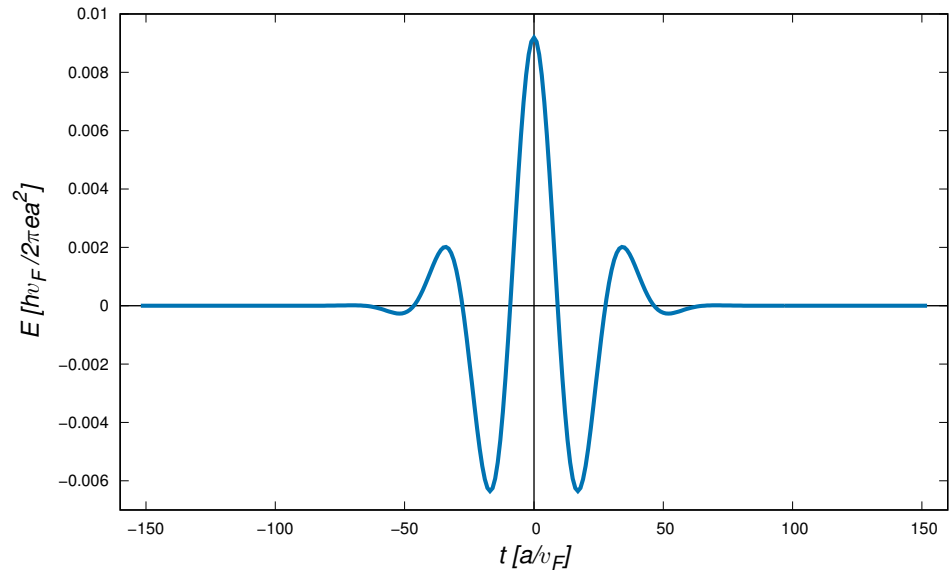}
\caption{Dependence of field strength on time in the model of the short
laser pulse (\protect\ref{4.1}). We used the natural graphene units for time
$[a/v_F]$ and field strength $[\hbar v_F/ea^2]$ ($a$ is the graphene lattice
constant).}
\label{F2}
\end{figure}

\begin{figure}[H]
\centering
\includegraphics[width=13 cm]{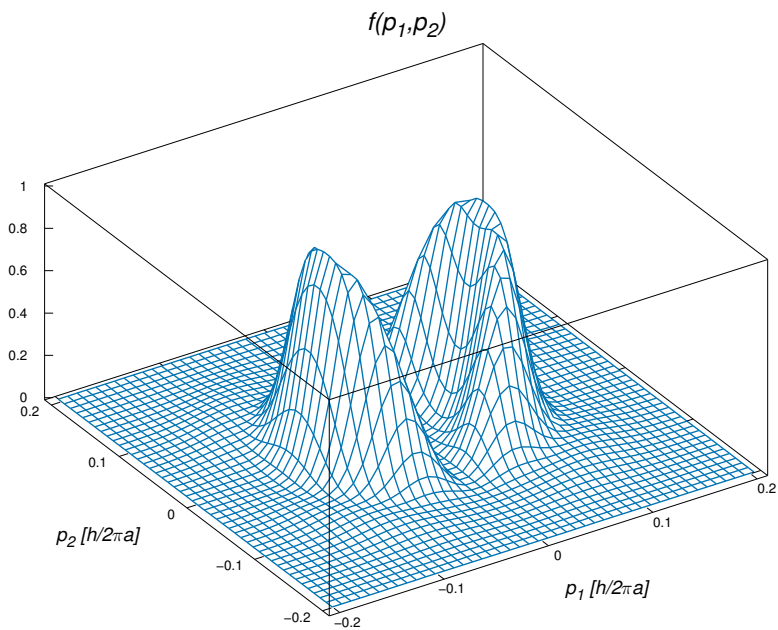}
\caption{Distribution function $f(\mathbf{k},t\rightarrow\infty)$ after the
end of the pulse calculated on the basis of KEs (\protect\ref{2.28}). The
values of the momentum components are given in units of $[ \hbar /a $].}
\label{F3}
\end{figure}

\begin{figure}[tbp]
\centering
\includegraphics[width=13 cm]{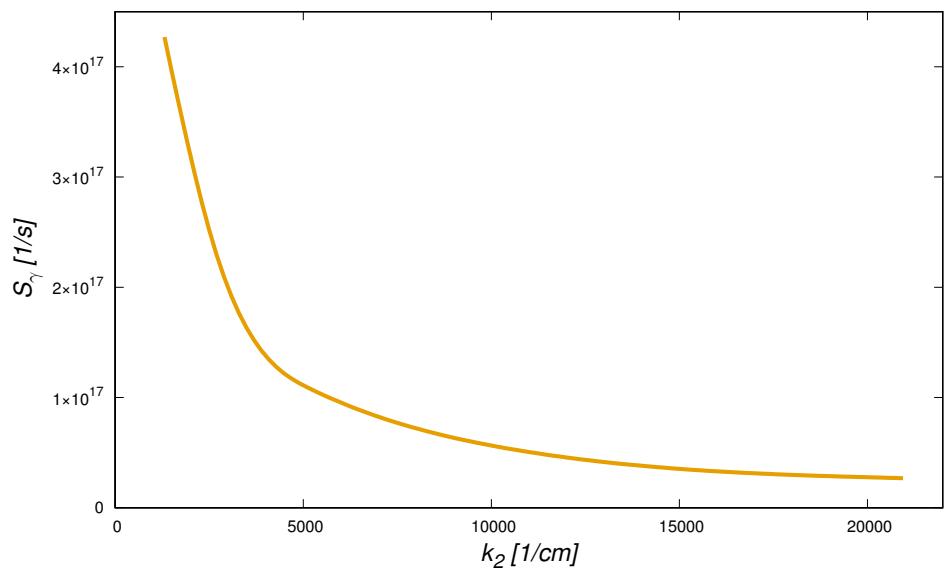}
\caption{$1/k$ - structure of the CI (\protect\ref{3.1}) of the annihilation
channel obtained as the result of direct numerical calculations $(k_{1}=0,\
t=100\ a/V)$}
\label{F4}
\end{figure}

\begin{figure}[H]
\centering
\includegraphics[width=13 cm]{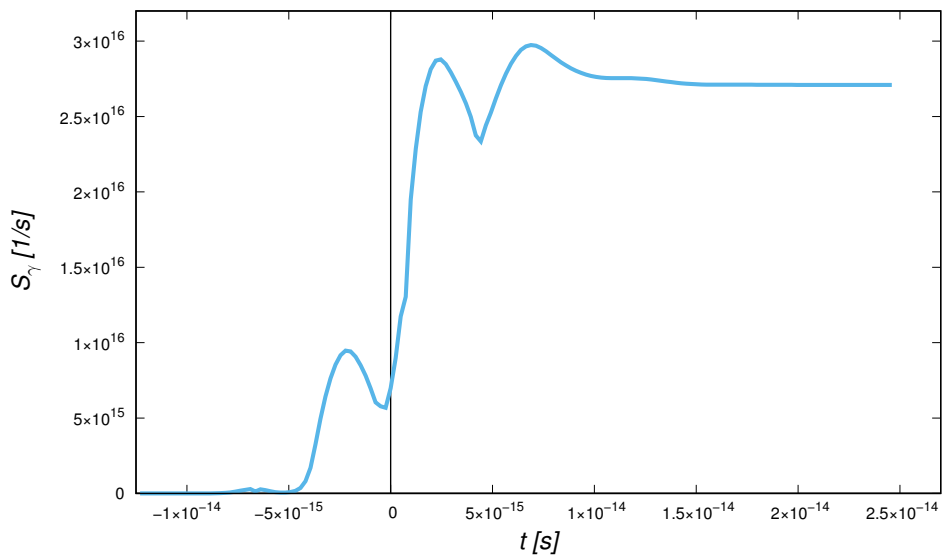}
\caption{Time dependence of the collision integral $S_{\protect\gamma}(%
\mathbf{k},t)$ }
\label{F5}
\end{figure}

$1/k$-dependence in the region of the central peak is reproduced also on the
qualitative level by virtue of numerical calculations directly of CI (\ref%
{3.1}) in the annihilation channel.

Specific of the computer calculations forces one to use the field model of a
short laser impulse (\ref{4.2}) (see Fig.\ref{F2} for the case of linear
polarization, $E_{0}=10^{6}\ {\mathrm{V}/cm},\ \omega =2\pi \times 10^{14}\
\mathrm{Hz},\ \sigma =3$). The distribution function of the \textrm{eh}%
-subsystem calculated on the basis of basic KE (\ref{2.28}) is depicted for
the point of time $t=0$ on the Fig.\ref{F3}. The valley $p_{1}=0$ between
two symmetrical fragments of the distribution function is a consequence of
the structure of amplitude (\ref{2.10}) ($E_{\mathrm{ext}}^{1}(t)=0$ by
virtue of the chosen polarization of the external field). At last, Fig.\ref%
{F4} shows $1/k$ dependence in projection on the axis of $k_{2}$ calculated
on the basis of CI (\ref{3.1}) of the annihilation channel.

(6) As it was showed in Sect.\ref{V}, the considered above CI in the model
of periodical field (\ref{4.3}) do not contain time independent components.
This fact can be explained by the absence of a definite asymptotic limit of
field model (\ref{4.3}) at $t\rightarrow \infty $ \cite{Blaschke:2013}. In
the general case CI can have non-zero asymptotic at $t\rightarrow \infty $
if the back reaction of the system is not taken into account. Fig.\ref{F5}
demonstrates such situation in the model of a short laser pulse (\ref{4.1}).
We believe that a special consideration of such behavior is would be quite
important.

\section{Conclusion}

\label{VI}

In the present work we have considered some physical processes which are
possible in the graphene under action of a strong time dependent electric
field. To this end nonperturbative methods (see \cite%
{Gitman:1977,Fradkin:1981,Fradkin:1991} and relevant Refs. given in the text
of the article) of strong field QED with unstable vacuum (in particular, the
Dirac model of the graphene) were used in combination with kinetic
description of radiation from the electron-positron plasma created from the
vacuum under an action of a strong time dependent electric field developed
in Refs. \cite%
{Smolyansky:2020,Smolyansky:2020a,Blaschke:2011,Smolyansky:2019b,Otto:2017}.
We would like to emphasize the significant development of KE formalism and
the study of the structure of these equation presented in the work. This
formalism includes a nonperturbative basis oriented on the description of
excitations of the \textrm{eh}-plasma in the graphene under the action of
strong electric fields and a perturbative part described interaction with
the quantized electromagnetic field. Main concrete results obtained in the
work on the base of such generalized theory are presented in Subsect. \ref%
{Vc}. We stress that some of the predicted properties of the model under
consideration may be verified experimentally. First, is the issue of the
possible presence of even harmonics of the external field in the quantum
radiation spectrum. Another important property that can be tested is the
characteristic spectral composition anisotropy of the quantum radiation and
its direction in the graphene plane (emission in the external space of the
electromagnetic waves of the plasma oscillations is oriented perpendicular
to the graphene plane). Some other predictions are related to the long wave
features of this radiation. In this respect, we note that the developed
approach can be extended to a wider range of parameters of the external
field including into the consideration arbitrary polarizations. It is also
important to consider in a similar manner such processes as photoproduction
of the \textrm{eh} - plasma under the action of quantum electromagnetic
field, cascade processes, the Breit-Wheeler process (e.g., \cite{Titov:2016}%
) and so on. We recall that a radiation of quasiclassical plasma waves in
the graphene obtained in the frame work of the nonperturbative KE approach
had got sufficiently reliable experimental testing (e.g., \cite%
{Smolyansky:2020b} and having there references).

In the conclusion, it should be noted that the new extended formulation of
KE approach contains some model assumptions. Further calculations in the
frame work of the formulation may check validity of these assumptions. Such
checking would be important for applications of strong field QED in $3+1$
dimensions, where the situation is more complicated.

\funding{This research was funded by Russian Science Foundation (Grant no.
19-12-00042)}

\acknowledgments{V.V.D., A.D.P. and S.A.S. are grateful to A.V. Tarakanov
for discussions and assistance in preparing this manuscript.}

\conflictsofinterest{The authors declare no conflict of interest.}

\abbreviations{The following abbreviations are used in this manuscript:\\

\noindent
\begin{tabular}{@{}ll}
KE & kinetic equation\\
CI & collision integral\\
eh  & electron-hole\\
BBGKY & Bogoliubov–Born–Green–Kirkwood–Yvon
\end{tabular}}

\reftitle{References}






\end{document}